\documentclass[12pt]{article}
\usepackage{a4wide}
\usepackage{bbold}
\usepackage{amsmath,amssymb,bbm}
\usepackage{multirow}
\usepackage{comment}
\newsavebox{\uuunit}
\sbox{\uuunit}
    {\setlength{\unitlength}{0.825em}
     \begin{picture}(0.6,0.7)
        \thinlines
        \put(0,0){\line(1,0){0.5}}
        \put(0.15,0){\line(0,1){0.7}}
        \put(0.35,0){\line(0,1){0.8}}
       \multiput(0.3,0.8)(-0.04,-0.02){12}{\rule{0.5pt}{0.5pt}}
     \end {picture}}

\def\2{\frac12}
\def\4{\frac14}

\newcommand{\be}{\begin{equation}}
\newcommand{\ee}{\end{equation}}
\newcommand{\bea}{\begin{eqnarray}}
\newcommand{\eea}{\end{eqnarray}}

\def\e{\epsilon}

\begin{document}

\begin{titlepage}
\begin{center}


\vskip 1.5cm

{\Large \bf  Heterotic  Wrapping Rules}

\vskip 1cm

{\bf Eric A.~Bergshoeff\,$^1$ and Fabio Riccioni\,$^2$}

\vskip 25pt

{\em $^1$ \hskip -.1truecm Centre for Theoretical Physics,
University of Groningen, \\ Nijenborgh 4, 9747 AG Groningen, The
Netherlands \vskip 5pt }

{email: {\tt E.A.Bergshoeff@rug.nl}} \\

\vskip 15pt

{\em $^2$ \hskip -.1truecm
 INFN Sezione di Roma,  Dipartimento di Fisica, Universit\`a di Roma ``La Sapienza'',\\ Piazzale Aldo Moro 2, 00185 Roma, Italy
 \vskip 5pt }

{email: {\tt Fabio.Riccioni@roma1.infn.it}} \\

\end{center}

\vskip 0.5cm

\begin{center} {\bf ABSTRACT}\\[3ex]
\end{center}

We show that the same wrapping rules that have been derived for the branes of IIA and IIB string theory also apply to the
branes of the toroidally compactified heterotic string theory. Moreover, we show that applying these wrapping rules to the IIA theory compactified over K3 is consistent with the well-known duality between the heterotic string theory  compactified over $T^4$ and the IIA string theory compactified over K3.

We derive a simple rule that relates, in any dimension, the T-duality representation of the branes of the toroidally compactified heterotic theory to the relevant
R-symmetry representation of the central charges in the supersymmetry algebra. We show that, in the general case, the degeneracy of the BPS
conditions of the heterotic branes is twice as large as that of the branes of IIA and IIB string theory.

\end{titlepage}

\newpage
\setcounter{page}{1} \tableofcontents

\newpage

\setcounter{page}{1} \numberwithin{equation}{section}

\section{Introduction}

It is well-known what the field content is of the maximal and half-maximal supergravity theories in different dimensions as long as one restricts oneself to the metric and  the $p$-form potentials that describe the physical states of the theory \cite{book}.
 These ``standard'' supergravity theories are complete in the sense that they are related
to each other by toroidal compactification\,\footnote{We do not consider here the gauged supergravity theories.}.
It has been appreciated for some time that the standard supergravity theories can be extended
by including the dual potentials. A special class is formed by the $(D-2)$-form potentials that are dual to the $0$-form potentials, or scalars, in the sense that the duality relations do not imply that the number of $(D-2)$-form potentials equals the number of scalars. Moreover, it has been realized that the
supergravity theories can be extended with $(D-1)$-form or deformation potentials that are dual to
mass parameters and $D$-form or top-form potentials, with identically vanishing curvatures, that do not describe any physical degree of freedom. All these potentials are relevant due to their coupling to branes. By definition, the ``standard'' potentials of supergravity couple to branes with 3 or more transverse directions. The other ``non-standard'' potentials couple to branes with 2 or less transverse directions. These branes are called defect branes (2 transverse directions),
domain walls (1 transverse direction) and space-filling branes (no transverse directions).
The non-standard branes are different from the standard ones in several respects.
First of all, to obtain finite-energy configurations, one must consider a collection of non-standard branes
in conjunction with orientifolds. Secondly, unlike the potentials of standard maximal supergravity, not all components of the T-duality representation to which the non-standard potentials belong correspond to {\sl half-supersymmetric} branes. For the
case of maximal supergravity we have introduced an elegant so-called ``light-cone rule''
that identifies which components of the representation correspond to elementary half-supersymmetric branes. In general, the remaining components correspond to (threshold or non-threshold) bound states of
branes or to branes with less supersymmetry \cite{Lu:1997bg}.

The most striking difference between standard and non-standard supergravity, however, is the fact that
{\sl non-standard supergravity is not complete} in the sense that the non-standard potentials in different dimensions are {\it not} related to each other by toroidal compactification. This has the important consequence that the
half-supersymmetric branes of non-standard supergravity, upon toroidal reduction, do {\it not} organize themselves into representations of T-duality, which is in conflict with string theory. To fill up complete T-duality representations one needs the emergence of extra branes upon compactification. How these extra  branes
should emerge is at the heart of the long sought for geometry underlying string theory. In our previous work we have advocated a particular approach to this crucial issue. Our starting point is to consider the inclusion of extra mixed-symmetry fields to the non-standard supergravity multiplet\,\footnote{In principle one can derive which mixed-symmetry multiplets can be added to a given supergravity multiplet by requiring linearized supersymmetry. In practice, they are derived by analyzing the spectrum of the very-extended Kac-Moody algebra ${\rm E}_{11}$, see below.}.
A particularly important example of such a mixed-symmetry field is the dual graviton. It is well-known that, using standard local field theory,
mixed-symmetry fields such as the dual graviton can only be defined, consistent with supersymmetry,
at the linearized level \cite{Bekaert:2002uh}. It is
an open problem whether and how this can be extended to the non-linear level. Our interest in these
mixed-symmetry fields is motivated by the fact that they contain precisely the information about how many extra branes should be produced upon dimensional reduction to restore T-duality. In earlier work we have  argued that these
mixed-symmetry fields couple to a new class of branes which are generalizations of the Kaluza-Klein monopole. However, this is not needed for the present purpose. Alternatively, one can argue that the mixed-symmetry fields
encode information about the geometry underlying string theory. Indeed, we will point out below that
they give rise to interesting brane wrapping rules that suggest a stringy generalization of the usual geometry.
Remarkably, precisely the same mixed-symmetry fields that are needed to restore T-duality occur, for the case of maximal supergravity, in the spectrum of the very extended Kac-Moody algebra ${\rm E}_{11}$. This algebra has been advocated as the one
underlying the symmetries of M-theory \cite{West:2001as}.
This approach suggests an extension of spacetime with extra coordinates and in this way to go
beyond standard local field theory \cite{West:2003fc}. In this work we will remain within local field theory but we will make use,
 at several places, of this relation with ${\rm E}_{11}$
and, in the case of half-maximal supergravity,  with other very extended Kac-Moody algebras as well.

In our earlier work, we pointed out that in the case of maximal supergravity the effect of the extra branes, following from the mixed-symmetry fields upon toroidal reduction, is not only that the
branes organize themselves into T-duality multiplets but, furthermore, that they can be understood as the result of certain  {\sl brane wrapping rules} \cite{Bergshoeff:2011ee}\,\footnote{In lower-dimensional supergravities we also find  T-duality multiplets of branes of which {\it none} component
follows from the reduction of a ten-dimensional brane. These branes are less well understood in the sense that one cannot formulate wrapping rules for them.
}. The explicit form of these wrapping rules will be given later in the paper. These rules  differ
from the naive brane wrapping rules corresponding to standard geometry.  The wrapping rules corresponding to standard geometry prescribe that {\sl any} brane, whether wrapped or unwrapped, leads to a single brane in one dimension lower:
    \begin{eqnarray}\label{standardgeometry}
 & & {\rm any\ brane} \ \ \ \left\{ \begin{array}{l}
{\rm wrapped} \ \ \ \ \rightarrow\  \ \ {\rm undoubled}\\
{\rm unwrapped} \ \ \rightarrow \ \ {\rm undoubled}
 \end{array} \right. \,.
 \end{eqnarray}
The
new wrapping rules, given later in this work, prescribe that, in certain cases, the brane can double when wrapped or un-wrapped. These
new wrapping rules can be said to define a stringy generalisation of standard geometry\,\footnote{An alternative scenario is that one
works with standard
geometry but supplemented with a new kind of branes, such as the generalized Kaluza-Klein monopoles mentioned above.}. The purpose of this work is to extend this analysis to the case of half-maximal supersymmetry. Remarkably, we find that {\it the same} wrapping rules that we found in the maximal case can also be used to
reduce the branes of the 10D heterotic string. Moreover, applying these same wrapping rules  to the
K3 orbifold we reproduce the well-known duality, at the level of supersymmetric branes\,\footnote{In this
paper we will also discuss the duality at the level of the complete spectrum of the very extended Kac-Moody algebra, i.e.~including the mixed-symmetry fields.  In order to do this we need to define a $\mathbb{Z}_2$ truncation on the IIA side. How to precisely define such a $\mathbb{Z}_2$ truncation at the level of the complete spectrum is not yet clear to us. This issue will be discussed further in the paper.},
between the heterotic string compactified over $T^4$ and the IIA string compactified over K3. The fact that  the same wrapping rules can be used suggests that different geometries, such as the torus and the K3 orbifold, allow for the same stringy generalization.

In this work we will also clarify the relation between the number of
half-supersymmetric branes and the BPS conditions that they satisfy, which
are related to the central charges of the supersymmetry algebra with 16
supercharges. It is well-known that in the case of the standard branes of
maximal supergravity
there is a 1-1 relation between the half-supersymmetric branes and the
central charges: for each central charge, and its dual, there is a single
half-supersymmetric brane \cite{de Azcarraga:1989gm,Townsend:1997wg}. For
the non-standard branes the situation is more subtle due to the fact that
degeneracies occur: one central charge, or BPS condition, may correspond
to more half-supersymmetric branes. For the defect branes of maximal
supergravity we found that the degeneracy in each dimension is two: each
defect-brane and its S-dual satisfy the same BPS condition.
The degeneracies in the case of the domain walls of maximal supergravity
have recently been investigated \cite{Bergshoeff:2012pm}.
We will show that the degeneracies that occur in the case of the heterotic
branes are always twice the ones we found for the branes of maximal
supergravity. This includes the standard heterotic branes which have a
degeneracy 2 instead of 1 like the branes of maximal supergravity.

In this work we will give, at different places, a few rules which are very useful for
several counting purposes. For the convenience of the reader we summarize them below:
\bigskip

\noindent {\bf light-cone rule}\,: this rule prescribes which components of the T-duality representation of a  $p$-form potential correspond to a half-supersymmetric brane. The rule is given in Subsection 2.1
\vskip .2truecm

\noindent {\bf restricted reduction rule}\,: this rule explains which components of a mixed-symmetry field, upon toroidal reduction, gives rise to a potential in lower dimensions  corresponding to a half-supersymmetric brane. It is given in Subsection 3.1
\vskip .2truecm

\noindent {\bf heterotic truncation rule}\,: this rule shows how the  branes of the toroidally compactified heterotic string theory
can be obtained by truncating the branes of toroidally compactified IIA or IIB string theory, see Subsection  3.2
\vskip .2truecm

\noindent {\bf central charge rule}\,: this rule relates, in any dimension, the T-duality representation of the branes of the toroidally compactified heterotic theory to the relevant
R-symmetry representation of the central charges in the supersymmetry algebra with 16 supercharges. The rule can be found in Section  5
\bigskip

We conclude with summarizing the outline of this work. In Section 2 we first classify the  half-supersymmetric branes of the toroidally compactified heterotic string theory using the light-cone rule mentioned above. We then show that the branes in lower dimensions, that occur in a T-duality representation
that contains at least one brane that follows from the reduction of a 10D brane, can be obtained
by introducing a set of heterotic wrapping rules which we will specify. In Section 3 we show how the branes of the toroidally compactified heterotic string theory
can be obtained by truncating the branes of toroidally compactified IIA or IIB string theory. We will also discuss issues that arise when one tries to extend this so-called
heterotic truncation of the $p$-form potentials that couple to supersymmetric branes to the full spectrum of fields, including the mixed-symmetry fields. In Section 4 we show that applying the wrapping rules  of the maximal theories to the IIA string theory compactified over K3
leads precisely, at the level of the supersymmetric branes, to the well-known duality between the heterotic theory on $T^4$ and the IIA theory on K3. We also discuss  the wrapping rules of the  IIB theory on K3. Next, in Section 5 we discuss the relation between the heterotic branes and the central charges in the supersymmetry algebra with 16 supercharges. In particular,
we show that, in the general case, the degeneracy of the BPS
conditions of the heterotic branes is twice as large as that of the branes of IIA and IIB string theory. Finally, in Section 6 we present our conclusions. We have added two appendices. In appendix A we discuss  some properties of the $\text{SO}(8,8+n)^{+++}$ very extended Kac-Moody algebra. In particular, we discuss in this appendix the definition of the real roots in the non-split case, i.e.~$n\ne -1,0,1$. In appendix B we discuss the truncation of the IIB theory to the closed sector of the Type I string theory.

\section{Heterotic Branes and Wrapping Rules}

This section contains two Subsections. In the first Subsection we will  determine  the half-supersymmetric branes
of the heterotic string theory compactified on the $d$-dimensional
torus $T^d$. In the next Subsection we will define wrapping rules for these
heterotic branes.

We remind the reader that
the low-energy effective action of the toroidally compactified heterotic theory  is half-maximal
supergravity coupled to vector multiplets. Generically, Wilson lines
break the gauge group (either $\text{SO}(32)$ or $\text{E}_8 \times
\text{E}_8$) to $\text{U}(1)^{16}$. Including also the vectors
arising from the metric and the NS-NS 2-form,  this gives a total of
$16 +2d $ abelian vectors. These vectors transform in the
fundamental representation of the T-duality symmetry group
$\text{SO}(d, 16+d)$, while the scalars parametrise the coset
manifold  $\text{SO}(d, 16+d)/ [ \text{SO}(d)\times
\text{SO}(16+d)]$. Together with gravity, the NS-NS 2-form and the
dilaton, this is the bosonic content of  the $D=10-d$ dimensional
gravity multiplet plus $16+d$ abelian vector multiplets. Indeed, in
each dimension $D=10-d$ the bosonic field content of the gravity
multiplet is
\begin{equation}
e_\mu{}^a \quad B_2 \quad d\times  B_1 \quad \phi \quad ,
\end{equation}
where $B_2$ is a 2-form and $B_1$ are vector fields, while
the bosonic field content of the vector multiplet is given by
\begin{equation}
B_1 \quad d\times \phi
\quad .
\end{equation}
The 2-form of the gravity multiplet is dualised to a vector in $D=5$
and to a scalar in $D=4$, parametrising the manifold
$\text{SL}(2,\mathbb{R})/ \text{SO}(2)$ together with the dilaton. In $D=3$ all the vectors are dualised to scalars, and the resulting scalars (including the dilaton) parametrise the manifold $\text{SO}(8,24)/[ \text{SO}(8) \times \text{SO}(24)]$.

\subsection{Half-supersymmetric heterotic branes}

In classifying the heterotic branes it is natural to label these
branes, and the corresponding fields, according to the way the
tension scales with the dilaton in the string frame. The 2-form and
1-forms are thus called fundamental fields in the sense that they
couple to fundamental objects whose tension is independent of the
dilaton. Using a notation where the dependence of the brane tension
$T$ on the $D$-dimensional dilation $\phi$ is specified by a number
$\alpha$
\begin{equation}
T\ \sim \e^{\alpha\phi}
\end{equation}
it means that $\alpha=0$ for these fields. Similarly, the dual
solitonic objects ($(D-5)$-branes and $(D-4)$-branes) have
$\alpha=-2$, and are electrically charged under the Poincare duals
of the 1-forms and 2-form. In $D>4$, the 1-forms, the 2-form and their duals are all standard potentials, i.e. they are associated to branes with at least three transverse directions.

Besides the standard potentials, the following non-standard potentials occur:

\begin{enumerate}
\item  $(D-2)$-form potentials. They satisfy duality relations with the scalars of the supergravity coset models.
They are special in the sense that  the number of such potentials is
not equal to the number of coset scalars.  Instead, they satisfy
extra curvature constraints. These potentials couple to branes with
two transverse directions, i.e.~defect branes.

\item  $(D-1)$-form potentials or deformation potentials. These potentials are the duals of mass parameters and do not describe any physical degrees of freedom.
They couple to branes with one transverse direction,  i.e.~domain
walls.

\item $D$-form potentials or top-form potentials.  These potentials have an identically vanishing curvature and couple to space-filling branes.
    \end{enumerate}

These potentials, as well as the Poincare duals of the 1-forms and
2-form, were not included in the multiplets  above, because they are
either dual to the potentials describing physical degrees of freedom  or they do not carry any
on-shell degree of freedom, like the $(D-1)$-forms and $D$-forms.
Nevertheless, they can be introduced in the supersymmetry algebra, and they give
important information about the heterotic branes. They transform as
representations of the global symmetry  group $\text{SO}(d,d+n)$ where $n$ is
the number of vector multiplets in 10 dimensions. These
representations can be determined by requiring the closure of the
supersymmetry algebra.  In \cite{Schnakenburg:2004vd} it was shown that these theories in any dimension are associated to the very-extended Kac-Moody algebras $\text{SO}(8,8+n)^{+++}$, and as a consequence of this the representations of the forms
 can also be obtained by
analysing the roots of these algebras
\cite{Bergshoeff:2007vb}. Among all the
forms that one obtains by this analysis, we are only interested in
those forms in $D$ dimensions that are associated to {\sl
half-supersymmetric} branes. In the maximal case, all the
half-supersymmetric branes have been obtained using two different
methods. One method consists in writing down the Wess-Zumino (WZ)
term for a brane electrically charged under the corresponding
potential \cite{Bergshoeff:2010xc,Bergshoeff:2011zk,Bergshoeff:2011qk,Bergshoeff:2012ex}. A brane is supersymmetric if the worldvolume fields that
one has to introduce to make the WZ term gauge invariant fit within
the bosonic sector of a half-supersymmetric multiplet. The other
method consists in analysing the $\text{E}_{8}^{+++}$ roots
associated to the gauge potentials. If the root is real, then the
corresponding potential is associated to a half-supersymmetric brane
\cite{Houart:2011sk,Kleinschmidt:2011vu}. These two methods give the
same answer.

In this paper we want to  make a similar analysis for the
half-maximal theories. The extension of both the Kac-Moody and the
WZ methods to the half-maximal case is non-trivial. In the
Kac-Moody approach this is related to the fact that in the
non-split case, i.e.~$n\ne 0,1$, the notion of real root has to be
refined. We will show
in Appendix   \ref{appendix1} that in going from the split case $n=0,1$  to the
non-split case (generic $n$) the representations of the forms are
naturally extended from $\text{SO}(d,d)$ to $\text{SO}(d,d+n)$ but the
associated real roots, or half-supersymmetric branes, are given by a
so-called ``light-cone rule'' which is given below. Similarly, the
WZ analysis has to be supplemented by an analysis of
certain supersymmetry cancellations between the Nambu-Goto and
WZ terms, as explained below.

Independent of whether we use the Kac-Moody or WZ method
the outcome of our analysis is that  all the fields in the
$D$-dimensional heterotic theory that are associated to half-supersymmetric branes
are the ones given in Table \ref{halfsusybranefields}. In this table we have also
specified the corresponding value of $\alpha$. Not all
components of these fields correspond to supersymmetric branes. We
find that the precise number of half-supersymmetric heterotic branes
is derived from the given representations using the following
so-called light-cone rule: \vskip .3truecm

\noindent {\bf light-cone rule}\,:\ \ \   Given a potential in a representation  of the duality group
$\text{SO}(d,d+n)$ we split the $2d+n$ duality indices into $2d$ `lightlike' indices  $i \ \pm\ (i=1,...,d)$
and the remaining $n$ `spacelike' indices.
A given component of the potential couples to a
half-supersymmetric brane if one of the following situations apply:

\begin{description}

\item{1.} {\sl anti-symmetric tensor representations}\,:\
the antisymmetric indices
are of the form $i\pm j\pm k \pm \dots $ with $i\,,j\,,k\,,\dots  $ all different.

\item{2.} {\sl mixed-symmetry representations}\,:  We only give this second rule for a
potential \\ $\phi_{A_1\dots A_m,B_1\dots B_n}\ (m>n)$ in a representation corresponding to a 2-column Young ta\-bleaux of heights $m$ and $n$
\footnote{The rule has a natural generalization to a multi-column Young tableaux.}. On top of the previous rule the following  additional rule applies:
each of the anti-symmetric $B$  indices in $\phi_{A_1\dots A_m,B_1\dots B_n}$
has to be parallel to one of the antisymmetric $A$ indices.

\end{description}
\vskip .1truecm

\begin{table}\small
\hskip -1truecm
\begin{tabular}{|c|c|}
\hline \rule[-1mm]{0mm}{6mm} $\alpha$ & fields \\
\hline
\hline \rule[-1mm]{0mm}{6mm}
$0$ & $B_{1,A} \quad B_2 \ (\text{H})$\\[.1truecm]\hline \rule[-1mm]{0mm}{6mm}
$-2$ & $D_{D-4}\ (\text{H})\quad  D_{D-3,A}\ (\text{H})\quad D_{D-2,A_1 A_2}\ (\text{H})\quad D_{D-1,A_1 A_2 A_3}\ (\text{H})\quad  D_{D,A_1 A_2 A_3 A_4}\ (\text{H})$\\[.1truecm]
\hline \rule[-1mm]{0mm}{6mm}
$-4$ &
 $F_{D-1,A_1\dots A_{d-3}}\ (\text{T,V,V,H,H})\quad F_{D,A,B_1\dots B_{d-3}}\ (\text{V,V,V,H}) \quad
F_{D-2, A_1\dots A_{d-6}}\ (\text{H})\quad  F_{D-1,A,B_1\dots B_{d-6}}\ (\text{H})$
 \\[.1truecm]
\hline \rule[-1mm]{0mm}{6mm}
$-6$ & $D=4$\,: $H_{4,A_1 A_2 A_3 A_4}\ (\text{H}) \qquad D=3$\,: $H_{2,A_1 A_2 A_3 }\ (\text{H}) \quad H_{3,A,B_1\dots B_5}\ (\text{H})$\\[.1truecm]
\hline \rule[-1mm]{0mm}{6mm}
$-8$ &  $D=3$\,: $J_2\ (\text{H}) \quad J_{3,A,B_1 ...B_4  }\ (\text{H})$\\[.1truecm]
\hline \rule[-1mm]{0mm}{6mm}
$-10$ & $ D=3$\,: $L_{3, A_1 ...A_4} \ (\text{H})$\\[.1truecm]
\hline
\end{tabular}
  \caption{\sl \footnotesize Universal $\text{SO}(d,d+n)$  representations for all half-supersymmetric heterotic branes of the half-maximal supergravity theory in $D$ dimensions that follows from the ten-dimensional ${\cal N}=1$ supergravity theory with $n$ vector multiplets.
  Capital indices $A$ refer to vector indices of the  group $\text{SO}(d,d+n)$. Repeated vector indices form anti-symmetric tensor representations. A comma between the vector indices indicates a mixed-symmetry representation. The worldvolume content (except for the $B_{1,A}$ fields) is indicated, starting from the highest possible dimension, between brackets with H,V,T indicating a Hypermultiplet, Vector  multiplet and Tensor multiplet, respectively. As it can be deduced from the table, we denote with different letters the fields corresponding to different values of $\alpha$. For more explanations, see the text.
  \label{halfsusybranefields}}
\end{table}

In the maximal case, the same light-cone rule was shown to occur in
terms of representations of the T-duality group
\cite{Bergshoeff:2011zk}. In that case,  the WZ method gives the
criterion that a potential can be associated to a
half-supersymmetric brane if the corresponding
gauge-invariant Wess-Zumino term requires the introduction of
world-volume fields that fit within the bosonic sector of a
supermultiplet with 16 supercharges. In this case we should require
that the corresponding world-volume fields describe the bosonic
sector of a multiplet with 8 supercharges.

We will now show how the Wess-Zumino method leads to the
classification of the half-supersymmetric heterotic branes, as given
by Table \ref{halfsusybranefields} supplemented  with the light-cone
rule. The bosonic content of the different multiplets with 8
supercharges, each of which describes 4+4 physical degrees of
freedom, are given by:
  \begin{itemize}
 \item $D=2,3,...,6$ hypermultiplet H with four scalars;
\item $D=4,5,6$ vector multiplet V with one vector plus two scalars ($D=4$), one scalar ($D=5$) or zero scalars ($D=6$);
\item $D=6$ tensor multiplet T with one self-dual tensor and one scalar.
\end{itemize}
Note that multiplets with 8 supercharges  only exist in $D\le 6$
dimensions. This is consistent with the fact that the fields in
Table \ref{halfsusybranefields} do not give rise to supersymmetric
branes according to the above light-cone rule when the rank of the
corresponding field is higher than 6. We should mention that in 2D
there are also multiplets with only chiral fermions or only chiral scalars,  that are singlets under
supersymmetry \cite{Bergshoeff:1989qh}. The reason is that, denoting with $x_{\rm L}$ and $x_{\rm R}$ the worldvolume light-cone coordinates in two dimensions, if one has supersymmetry in the left sector, then any field which only depends on $x_{\rm R}$ is automatically a singlet under supersymmetry.
Two relevant examples of these are the `heterotic fermions' that
play an important role in the construction of a gauge-invariant
worldvolume action of the heterotic string \cite{Atick:1985iy}, as well as the right sector of the transverse scalars.

We first show how the counting of  worldvolume degrees of freedom
works in $D=10$ dimensions. In ten dimensions we only have $B_2$ and
its dual $D_6$, together with $B_{1,A}$ and its dual $D_{7,A}$ which
are vectors of the compact group $\text{SO}(n)$. The fact that
$B_{1,A}$ cannot correspond to a supersymmetric brane can be seen by
looking at its supersymmetry variation. Indeed, this field only
transforms to the gaugino, and not to the gravitino. This implies
that one cannot write a $\kappa$-symmetric effective action: there
is no partial cancellation between the variation of  the WZ term and the variation of the
induced metric in the Nambu-Goto kinetic term. The fact that the dual
field $D_{7,A}$ does not lead to supersymmetric branes can be seen
by using the same argument or by simply noticing that a
seven-dimensional world-volume does not allow multiplets with 8
supercharges.
A special analysis is required for the worldvolume degrees of freedom of the 1-brane associated to the field $B_2$. This brane is the fundamental heterotic string, and we know that for such string  the left modes, that is the ones depending on $x_{\rm L}$, and the right modes, depending on $x_{\rm R}$, are different. In particular, only the left modes are supersymmetric, which means that out of the eight transverse scalars, only the part depending on $x_L$ fits within a supermultiplet. This counts as four degrees of freedom, that together with the left spacetime fermions form a hypermultiplet in two dimensions. The right modes are all singlets under supersymmetry, as already anticipated above.
There is a subtlety concerning the WZ term. Indeed, in the fermionic description of the heterotic theory, where one introduces 32 right-moving fermions producing a gauge symmetry that is either $\text{SO}(32)$ or $\text{E}_8 \times \text{E}_8$, there are no internal scalars and thus the WZ term is simply $B_2$, which is not gauge invariant.
However, there is an anomalous coupling between the heterotic fermions and
the 1-forms whose anomalous variation precisely cancels the gauge
transformation of the WZ term \cite{Atick:1985iy}.
In the bosonic description, instead, one introduces 16 internal bosonic right-moving coordinates $b_{0,A}$, and the WZ term is\,\footnote{We only
consider the general form of the WZ term and ignore the precise
values in front of the different terms. We also assume that
whatever can occur, does occur.}
\begin{equation}
B_2 + B_{1,A} {\cal F}_1^A\  , \label{WZB2}
 \end{equation} with ${\cal F}_{1,A}= d b_{0,A} + B_{1,A}$. These right-moving scalars, together with the right-moving transverse scalars, are singlets under supersymmetry.
Finally, for the solitonic 5-brane associated to $D_6$ one only gets four transverse scalars, corresponding again to a hypermultiplet.
Note that, since only embedding scalars are involved in the multiplets, no branes can
end on these objects.

We now proceed with the analysis of the  WZ terms in all dimensions.
First of all, for the 1-forms $B_{1,A}$ one has to rely on the
consideration of the  supersymmetry cancellations between the
Nambu-Goto and WZ terms to get the supersymmetric branes. This is
like in the ten-dimensional case. It turns out that only the {\sl
lightlike} directions of $\text{SO}(d,d+n)$ lead to fields that vary
under supersymmetry into  both the gravitino and the gaugino in such
a way that a cancellation between the WZ term and the Nambu-Goto
term can occur. We next consider the fundamental $B_2$ field.
The analysis here resembles the 10-dimensional one. In the bosonic description,
the WZ term is as in eq. \eqref{WZB2}, with the index $A$ now being an index of  $\text{SO}(d,d+n)$.
In this expression, the worldvolume field-strengths  are meant to satisfy duality relations. Splitting the T-duality directions into $2d$ lightlike directions and $n$ spacelike directions,  these duality relations are actually self-duality relations for the $n$ scalars in the spacelike directions, implying that these scalars are right-moving, while for the lightlike directions they give $d$ independent scalars, which split into $d$ left-moving and $d$ right-moving scalars. Only the left-moving part is supersymmetric. Similarly, the $8-d$ transverse scalars split into $8-d$ left-moving
and $8-d$ right-moving scalars. Together, we thus obtain $d+(8-d)=8$
left-moving scalars as in ten dimensions. This counts as four degrees of freedom, and  it corresponds to  a two-dimensional hypermultiplet. The right-moving lightlike scalars, the right-moving transverse scalars  and the right-moving spacelike scalars are singlets of supersymmetry. In the fermionic description, these latter scalars are replaced by internal fermions and the WZ term only contains the scalars along the lightlike directions, so that only the $\text{SO}(d,d)$ part of  the $\text{SO}(d,d+n)$ symmetry is manifest. This finishes our discussion of the fundamental fields in Table \ref{halfsusybranefields}.

We next  consider the solitonic $D$ fields in Table
\ref{halfsusybranefields}.  For $D_{D-4}$ we have just 4 transverse
scalars, that is one hypermultiplet. For $D_{D-3, A}$ one
has the WZ term
  \begin{equation}
D_{D-3,A} + D_{D-4} {\cal F}_{1,A}
\end{equation}
giving one scalar plus  three transverse scalars, that is one
hypermultiplet. Similarly, all the other $D$ fields give rise to one
worldvolume hypermultiplet. In all these cases, as well as all the cases below, the WZ argument has
to be supplemented with the requirement that there is a cancellation
between the Nambu-Goto kinetic term and the WZ term. These two
requirements together lead to the light-cone rule formulated above.

We now consider the $F$ fields. The first case in which an $F$ field appears is for $d=3$ or $D=7$, in which case one gets a 6-form which is a singlet of $\text{SO}(3,3+n)$. This is the first $F$ field in Table \ref{halfsusybranefields}. The WZ term is given by\,\footnote{In general we denote with ${\cal H}_n$ the field-strengths of the $\alpha=-4$ worldvolume form fields $d_{n-1}$. The notation is taken from that used in the maximal case, see Table 3 in \cite{Bergshoeff:2012ex}.}
  \begin{equation}
F_6 + D_3 {\cal H}_3
\quad ,
\end{equation}
which describes a self-dual tensor and a transverse scalar, that is a tensor multiplet in the six-dimensional worldvolume.
In 6D the field $F_{D-1, A_1 ...A_{d-3}}$ couples via the WZ term
  \begin{equation}
F_{5,A} + D_{3,A} {\cal H}_2
\end{equation}
giving a  vector multiplet. In 5D it couples via the WZ term
  \begin{equation}
F_{4, A_1 A_2} + D_{2, [A_1} {\cal H}_{2, A_2]} + D_{3, A_1 A_2 } {\cal H}_1\,,
\end{equation}
giving again a vector multiplet. In 4D it couples via
\begin{equation}
F_{3, A_1 A_2 A_3}+ D_{2, [A_1 A_2} {\cal H}_{1, A_3 ]} \end{equation}
giving a hypermultiplet. Finally, in 3D it couples via
\begin{equation}
F_{2, A_1 ...A_4} + D_{1, [ A_1 A_2 } {\cal H}_{1, A_3 A_4 ]}
\end{equation}
giving a two-dimensional hypermulitplet.

We next consider the second $F$ field in Table \ref{halfsusybranefields}, i.e.~the $F_{D, A, B_1 ...B_{d-3}}$ term. In $D=6$ this gives
  \begin{equation}
  F_{6, A,B} + F_{5, (A} {\cal F}_{1, B)}
\quad,
\end{equation}
giving one vector (remember that according to the light-cone rule the index $B$ has to be the same as
the index $A$), that is a vector multiplet in six dimensions. In
$D=5$ one gets
  \begin{equation}
F_{5, A, B_1 B_2} + F_{4, B_1 B_1 } {\cal F}_{1, A} + D_{3 , B_1 B_2 } {\cal H}_{2, A}
\end{equation}
giving one vector and one scalar, i.e.~a vector multiplet in five dimensions. In $D=4$ one gets
  \begin{equation}
F_{4, A, B_1 B_2 B_3} + F_{3, B_1 B_2 B_3} {\cal F}_{1, A}+ D_{3, B_1 B_2 B_3} {\cal H}_{1,A} + D_{2, [B_1 B_2} {\cal H}_{2, B_3]A} \quad ,
\end{equation}
giving one vector (because of self-duality) plus two scalars, i.e.~a vector multiplet in four dimensions.
Finally, in $D=3$ one gets
  \begin{equation}
F_{3, A, B_1 B_2 B_3 B_4}  + F_{2, B_1 ...B_4} {\cal F}_{1, A} + D_{2, [B_1 B_2 B_3} {\cal H}_{1, B_4 ], A} \quad ,
\end{equation}
giving four scalars, i.e.~a hypermultiplet in three dimensions.

We now consider the third $F$ field in Table
\ref{halfsusybranefields}. We only consider the four dimensional
case (the 3D case corresponds to a 0-brane), which gives
\begin{equation}
F_2 + D_{1, A} {\cal H}_1{}^A \quad .
\end{equation}
The analysis of the worldvolume degrees of freedom in this case is the same as for the fundamental heterotic string.
Actually, the two strings are S-dual to each other. Indeed, in 4D, the $\text{SL}(2,\mathbb{R})$ symmetry identifies the T-duality representation of a $p$-form with a given weight $\alpha$ with the representation with weight $-\alpha-2p$.
This means that only the left-moving part of the transverse scalars and the left-moving part of the scalars in the lightlike directions collect to form a two-dimensional hypermultiplet, while the other scalars are singlets under supersymmetry.

Finally, the last $F$ field in Table \ref{halfsusybranefields}
only exists in three dimensions, where it gives
  \begin{equation}
F_{2, AB} + F_{1, (A} {\cal F}_{1, B)} + D_{1, (A | C} {\cal H}_{1}{}^C{}_{B)}
 \quad . \end{equation}
Given that the indices $A$ and $B$ have to be parallel, this gives one (L+R) scalar from the second term. Together with the single embedding scalar, this gives 1
degree of freedom in the left-moving sector. Using the fact that in the third term the
index $C$ can have 12 different values, this last term gives 6 (L+R)  scalars because of self-duality. In the left-moving (supersymmetric) sector this corresponds to
3 degrees of freedom. Therefore, in the left-moving part we find 4 degrees of freedom that form the bosonic sector of a two-dimensional hypermultiplet.

Among the $H$ fields,  the only one which is not related to the
other cases we already discussed by S-duality is the last field in
Table \ref{halfsusybranefields}. Indeed, the 4D S-duality discussed above implies that the first $H$ term is S-dual to $D_{4,A_1 A_2A_3A_4}$. The other two $H$ terms
only exist in 3D. In 3D, the T-duality symmetry $\text{SO}(7,7+n)$ is contained in $\text{SO}(8,8+n)$, which identifies  the T-duality representation of a $p$-form with a given weight $\alpha$ with the representation with weight $-\alpha-4p$. Hence, the second $H$ term is S-dual to $D_{2,A_1 A_2A_3}$ while the last $H$ term is S-dual to itself. This term leads to the following WZ term
  \begin{equation}
H_{3, A, B_1 ...B_5} + F_{2, [ B_1 ...B_4} {\cal H}_{1, B_5] A} \quad , \label{onlyHfield}
\end{equation}
giving four scalars or 1 hypermultiplet in three dimensions. Using the same 3D and 4D S-duality rules, one can show that all the fields
in the last two rows of Table \ref{halfsusybranefields} are S-dual to fields we previously considered.
This concludes our analysis of the branes of the
toroidally-compactified heterotic theory.

There is an additional
theory with half-maximal supersymmetry, namely the ${\cal N}=(2,0)$
six-dimensional theory describing gravity plus 21 tensor multiplets
arising from compactifying IIB string theory on K3
\cite{Townsend:1983xt}. The bosonic content of the relevant (2,0)
multiplets is
  \begin{eqnarray}
& &  {\rm gravity}: \qquad e_\mu{}^a \quad 5 \times B_2^+\,,
 \nonumber \\
& & {\rm tensor}: \qquad \ B_2^-  \quad 5 \times \phi \quad .
\end{eqnarray}
From the supergravity point of view one can consider in general $5+n$
tensor multiplets, so that  the 2-forms transform as vectors of
$\text{SO}(5,5+n)$, while the $5 \cdot (5+n)$  scalars parametrise the
manifold $\text{SO}(5,5+n)/[\text{SO}(5) \times \text{SO}(5+n)]$. This
theory is anomaly-free only if $n=16$.

From the analysis of the Kac-Moody algebra, we find that the fields
whose highest weights of $\text{SO}(5,5+n)$ representations are real
roots of the $\text{SO}(8,8+n)^{+++}$ Kac-Moody algebra are given by\,\footnote{Note
that there is no notion of $\alpha$ in this theory.}
 \begin{equation}
A_{2, A}\ (\text{H}) \qquad  A_{4, A_1 A_2} \ (\text{V})\qquad A_{6
, A, B_1 B_2} \ (\text{V})\quad , \label{fieldsinthe20theory}
\end{equation}
where we have already indicated the worldvolume content that results
from the following WZ analysis. For $A_{2,A }$ we get four
transverse scalars, i.e.~one hypermultiplet. For $A_{4,A_1 A_2}$ we
get a WZ term of the form
  \begin{equation}
A_{4, A_1 A_2}+ A_{2, [A_1 } {\cal F}_{2, A_2 ]} \quad ,
\end{equation}
giving one vector (because of self-duality) and two transverse
scalars, i.e. a vector multiplet in four dimensions. Finally, for
$A_{6, A, B_1 B_2}$ we get
 \begin{equation}
A_{6, A, B_1 B_2} + A_{4, B_1 B_2} {\cal F}_{2, A} \quad ,
\end{equation}
giving one vector, that is  a vector multiplet in six dimensions. In all cases,  the WZ analysis has to be supplemented
with the requirement that there is a cancellation
between the Nambu-Goto kinetic term and the WZ term. These two
requirements together lead to the light-cone rule, which gives the following
half-supersymmetric branes:
  \begin{eqnarray}
& & {\rm 1-branes}: \ \ 10\,, \nonumber \\
& & {\rm 3-branes}: \ \  40\,, \nonumber \\
& & {\rm 5-branes}: \ \ 80\,.  \label{branesiib}
\end{eqnarray}
In section 4 we will show how the same number of branes follow from
a set of `K3 wrapping rules' to be defined in that section.

\subsection{Heterotic wrapping rules}

From the analysis in the previous subsection we may determine the
number of half-supersymmetric heterotic branes in each dimension. We
first restrict to those fields that amongst their T-duality components
have at least one brane that follows from the reduction of a brane
of the 10D heterotic theory. The corresponding branes are the
fundamental and solitonic branes with $\alpha=0$ and $\alpha=-2$,
respectively. The precise numbers are given in Tables
\ref{Fwrapping} and \ref{Swrapping}.

Remarkably, like in the maximal case, the same numbers of branes are
reproduced if we assume that the fundamental heterotic branes
satisfy the fundamental wrapping rule \cite{Bergshoeff:2011ee}
    \begin{eqnarray}\label{Fbranewrapping}
 & & {\rm F} \ \ \ \left\{ \begin{array}{l}
{\rm wrapped} \ \ \ \ \rightarrow\  \ \ {\rm doubled}\\
{\rm unwrapped} \ \ \rightarrow \ \ {\rm undoubled} \,,
 \end{array} \right.
 \end{eqnarray}
and that  the solitonic heterotic branes satisfy the solitonic
wrapping rule
 \begin{eqnarray}\label{Sbranewrapping}
 & & {\rm S} \ \ \ \left\{ \begin{array}{l}
{\rm wrapped} \ \ \ \ \rightarrow\  \ \ {\rm undoubled}\\
{\rm unwrapped} \ \ \rightarrow \ \ {\rm doubled} \quad .
 \end{array} \right.
\end{eqnarray}

\vskip .2truecm

\begin{table}[h]
\begin{center}
\begin{tabular}{|c||c|c|c|c|c|c|c|c|}
\hline \rule[-1mm]{0mm}{6mm} Fp-brane &10D& 9D & 8D & 7D & 6D&5D&4D&3D\\
\hline \hline \rule[-1mm]{0mm}{6mm} 0&&2&4&6&8&10&12&14\\
\hline \rule[-1mm]{0mm}{6mm} 1&1&1&1&1&1&1&1&1\\
\hline
\end{tabular}
\caption{\sl \footnotesize Upon applying the fundamental wrapping rule (\ref{Fbranewrapping}) one obtains in each dimension
a singlet fundamental heterotic string and $2d$ fundamental heterotic 0-branes.
\label{Fwrapping}}
\end{center}
\end{table}

\begin{table}[h]
\begin{center}
\begin{tabular}{|c||c|c|c|c|c|c|c|c|}
\hline \rule[-1mm]{0mm}{6mm} Sp-brane &10D& 9D & 8D & 7D & 6D&5D&4D&3D\\
\hline \hline \rule[-1mm]{0mm}{6mm} 0&&&&&&1&12&84\\
\hline \rule[-1mm]{0mm}{6mm} 1&&&&&1&10&60&280\\
 \hline \rule[-1mm]{0mm}{6mm} 2&&&&1&8&40&160&560\\
 \hline \rule[-1mm]{0mm}{6mm} 3&&&1&6&24&80&240&\\
 \hline \rule[-1mm]{0mm}{6mm} 4&&1&4&12&32&80&&\\
 \hline \rule[-1mm]{0mm}{6mm} 5&$1$&$2$&$4$&$8$&$16$&&&\\
\hline
\end{tabular}
\caption{\sl \footnotesize Upon  applying the solitonic wrapping
rule (\ref{Sbranewrapping}) one obtains precisely the numbers of
half-supersymmetric heterotic solitons that follows from the WZ
analysis in this section.
 \label{Swrapping}}
\end{center}
\end{table}

\begin{table}[h]
\begin{center}
\begin{tabular}{|c||c|c|c|c|c|}
\hline \rule[-1mm]{0mm}{6mm} $\alpha=-4$ branes & 7D & 6D&5D&4D&3D\\
\hline \hline \rule[-1mm]{0mm}{6mm} 0&&&&&14\\
\hline \rule[-1mm]{0mm}{6mm} 1&&&&1&560+14\\
 \hline \rule[-1mm]{0mm}{6mm} 2&&&&160&2240\\
 \hline \rule[-1mm]{0mm}{6mm} 3& &&40&480&\\
 \hline \rule[-1mm]{0mm}{6mm} 4&&8&80&&\\
 \hline \rule[-1mm]{0mm}{6mm} 5&1&8&&&\\
\hline
\end{tabular}
\caption{\sl \footnotesize This table gives the $\alpha=-4$ $p$-branes of the heterotic theory in any dimensions. The
value of $p$ is indicated in the first column. These branes do not satisfy any specific wrapping rule.
 \label{Ffieldstable}}
\end{center}
\end{table}

\begin{table}[h]
\begin{center}
\begin{tabular}{|c||c|c|c|c|c|}
\hline \rule[-1mm]{0mm}{6mm} $\alpha<-4$ branes &4D&3D\\
\hline \hline \rule[-1mm]{0mm}{6mm} 0&&\\
\hline \rule[-1mm]{0mm}{6mm} 1&&$280_{-6} + 1_{-8}$\\
 \hline \rule[-1mm]{0mm}{6mm} 2&&$3360_{-6} + 2240_{-8} + 560_{-10}$\\
 \hline \rule[-1mm]{0mm}{6mm} 3& $240_{-6}$&\\
\hline
\end{tabular}
\caption{\sl \footnotesize This table gives the $p$-branes with $\alpha<-4$ that occur in 4D and 3D. The value of $p$ is indicated in the first column. The subscript denotes the value of $\alpha$.  These branes do not satisfy any specific wrapping rules.
 \label{loweralphatable}}
\end{center}
\end{table}

All the other branes that arise from the fields in Table \ref{halfsusybranefields} do not satisfy any specific wrapping rule. Correspondingly, in any dimensions these branes belong to T-duality multiplets that do not contain branes of the 10-dimensional theory. We give the number of these branes, as resulting from the analysis of this section and satisfying the light-cone rule, in Tables \ref{Ffieldstable} and \ref{loweralphatable}.
Taking all the results contained in Tables \ref{Fwrapping} to \ref{loweralphatable} together one can read off the full result.

In 4D there is an additional $\text{SL}(2,\mathbb{R})$ symmetry. One can see from the tables how the branes rearrange themselves in terms of this symmetry. The 4D fields that as representations of $\text{SO}(6,6+n)\times \text{SL}(2,\mathbb{R})$ have highest weights corresponding to real roots of $\text{SO}(8,8+n)^{+++}$ are
\begin{eqnarray}
& & {\rm 1-form}: \ \ A_{1,Aa}\,, \nonumber \\
& & {\rm 2-form}: \ \ A_{2,ab} \ \ A_{2, A_1 A_2}\,, \nonumber \\
& & {\rm 3-form}: \ \ A_{3, A_1 A_2 A_3}\,, \nonumber \\
& & {\rm 4-form}:\ \  A_{4, A_1 ...A_4 ab} \ \ A_{4 ,A ,B_1 B_2 B_3} \quad .
\end{eqnarray}
The indices $a,b=1,2$ are $\text{SL}(2,\mathbb{R})$ indices and for the field to correspond to a brane they have to be  parallel. The value of $\alpha$ is related to the rank $p$  of the form by
 \begin{equation}
\alpha= n_1 - n_2 - p \quad ,
\end{equation}
where $n_1$ and $n_2$ are the number of indices along the directions 1 and 2 of $\text{SL}(2,\mathbb{R})$.
The reader can check that these rules, together with the light-cone rule that selects the $\text{SO}(6,6+n)$ components, gives all the 4D branes in the tables.

In 3D, the $\text{SO}(7,7+n)$ symmetry gets enhanced to $\text{SO}(8,8+n)$. The 3-dimensional fields that as $\text{SO}(8,8+n)$ representations have highest weights associated to real roots of $\text{SO}(8,8+n)^{+++}$ are
  \begin{eqnarray}
& &  {\rm 1-form}: \ \ A_{1,\hat{A}_1 \hat{A}_2}\,, \nonumber \\
& & {\rm 2-form}: \ \ A_{2,\hat{A} \hat{B}} \ \ A_{2, \hat{A}_1 ... \hat{A}_4}\,, \nonumber \\
& & {\rm 3-form}: \ \ A_{3, \hat{A},  \hat{B}_1 ...\hat{B}_5}  \quad ,
\end{eqnarray}
where the hatted indices are $\text{SO}(8,8+n)$ vector indices and the symmetries of the indices are denoted as everywhere else in the paper (see the caption of Table  \ref{halfsusybranefields}). The reader can verify that applying the $\text{SO}(8,8+n)$   light-cone rule to these fields gives exactly the branes listed in Tables  \ref{Fwrapping} to \ref{loweralphatable}. The value of $\alpha$ is related to the rank $p$ of the form  by
\begin{equation}
\alpha =2( n_+ - n_- -p )
\quad ,
\end{equation}
where $n_+$ and $n_-$ are the number of indices along the lightlike directions $8+$ and $8-$. These are the indices that are not in the $\text{SO}(7,7+n)$ directions.

\section{Heterotic Truncation}

The aim of this section is to determine the half-supersymmetric  brane spectrum of the heterotic theory from a suitable truncation of the type II theories. In a separate appendix we will comment about the truncation of the Type IIB theory leading to the Type I superstring,
see appendix B.

It is known that the pure supergravity sector of the heterotic low-energy effective action in ten dimensions can be obtained from both the IIA and the IIB supergravity theories by well-defined truncations.
We want to show to what extent this can be generalised to the whole ten-dimensional spectrum of the corresponding Kac-Moody algebra, including the mixed-symmetry fields.
We assume that the ten-dimensional `gravity' sector of the heterotic theory is derived from the Kac-Moody algebra $\text{SO}(8,8)^{+++}$, while the `matter' sector corresponds to all the
ten-dimensional fields resulting from the Kac-Moody algebra $\text{SO}(8,8+n)^{+++}$ that are not contained in the gravity sector. In subsection 3.3  we will give an argument that justifies this assumption.
We will consider the IIA and IIB ten-dimensional spectrum resulting from the Kac-Moody algebra $\text{E}_8^{+++}$  and we will compare it to the ten-dimensional spectrum of the $\text{SO}(8,8)^{+++}$ algebra. We will show that the truncation is well-defined as long as one only considers the fields associated to the real roots. The picture is less clear when one considers all the fields in the spectrum. Given that the fields
corresponding to the real roots are those that give rise to branes after dimensional reduction, and given that the analysis of the previous section shows that no additional branes are introduced in the heterotic theory from the matter sector, this analysis shows that all the heterotic branes can be obtained by truncation. The fact that the  $\text{SO}(8,8)^{+++}$ algebra can be obtained as a suitable truncation of the $\text{E}_8^{+++}$ algebra was originally discussed in \cite{Schnakenburg:2004vd}, while the analogous relation for the over-extended algebras $\text{SO}(8,8)^{++}$ and $\text{E}_8^{++}$ was analysed in \cite{Kleinschmidt:2004dy}. In particular, in this last reference the authors show that $\text{SO}(8,8)^{++}$ is a subalgebra of  $\text{E}_8^{++}$ by identifying their common $\text{SO}(9,9)$ subalgebra.

Before discussing this `heterotic' truncation,
we will first review, in the next subsection,
the 10-dimensional origin of the $D$-dimensional potentials that couple to the half-supersymmetric branes
of toroidally compactified IIA/IIB string theory.

\subsection{Half-supersymmetric branes in IIA/IIB string theory}

The potentials of $D$-dimensional maximal supergravity that couple to supersymmetric branes can be
derived either by the Kac-Moody method \cite{Kleinschmidt:2011vu} or the Wess-Zumino method \cite{Bergshoeff:2011mh}.
The result for any dimension is listed  in Table
\ref{table2}.

\begin{table}\small
\begin{center}
\begin{tabular}{|c||c|}
\hline \rule[-1mm]{0mm}{6mm} $\alpha$ & tensors \\
\hline
\hline \rule[-1mm]{0mm}{6mm}
$0$ & $B_{1,A}\,, B_2$\\[.1truecm]
\hline \rule[-1mm]{0mm}{6mm}
$-1$ &  $C_{2n+1,a}\,, C_{2n,{\dot a}}$\\[.1truecm]
\hline \rule[-1mm]{0mm}{6mm}
$-2$ & $D_{D-4}\,, D_{D-3,A}\,,$\\[.1truecm]
&$D_{D-2,A_1 A_2}\,, D_{D-1,A_1 A_2 A_3}\,, D_{D,A_1 A_2 A_3 A_4}$\\[.1truecm]
\hline \rule[-1mm]{0mm}{6mm}
$-3$ &$E_{D-2, {\dot a}}\,, E_{D-1, A{\dot a}}\,, E_{D,A_1 A_2 {\dot a}}$\\[.1truecm]
\hline \rule[-1mm]{0mm}{6mm}
$-4$ & $F^+_{D,A_1\dots A_d}$\\[.1truecm]
\hline
\hline \rule[-1mm]{0mm}{6mm}
 & $F_{D-1,A_1\dots A_{d-3}}\,, F_{D,A,B_1\dots B_{d-3}}$\,, \\[.1truecm]
&$F_{D-2, A_1\dots A_{d-6}}\,, F_{D-1,A,B_1\dots B_{d-6}}$\\[.1truecm]
\hline \rule[-1mm]{0mm}{6mm}
$-5$ &  $G_{D,A_1\dots A_{d-4},{\dot a}}\,,$ \\[.1truecm]
&$G_{D-1,A_1\dots A_{d-6},a}\,, G_{D,A,B_1\dots B_{d-6},a}$\\[.1truecm]
\hline \rule[-1mm]{0mm}{6mm}
$-6$ & $D=4$\,: $H_{4,A_1 A_2 A_3 A_4}$\\[.1truecm]
&$D=3$\,: $H_{2,A_1 A_2 A_3 }\,,H_{3,A,B_1\dots B_5}$\\[.1truecm]
\hline
\end{tabular}
\end{center}
  \caption{\sl \footnotesize Universal T-duality representations for all half-supersymmetric branes of the maximal supergravity theories in $D$ dimensions. The fields above the double horizontal line contain amongst their T-duality components
  at least one brane that follows from the reduction of a brane of IIA/IIB string theory. On the other hand, the fields below the double horizontal line contain
  none brane that follows from the reduction of a brane of IIA/IIB string theory.
  Capital indices $A$ refer to vector indices of the T-duality group $\text{SO}(d,d)$ with $d=10-D$.
    The indices $a,\dot a$ refer to chiral and
  anti-chiral spinor indices.
   Repeated vector indices form anti-symmetric tensor representations.
  Tensor representations with both $A$ and $B$ indices, separated by a comma, refer to mixed-symmetry representations, see section 2. The fields in $D=4$ and $D=3$ that have $\alpha<-6 $ are not included in the table. They are related to the fields with higher $\alpha$ by $\alpha \rightarrow  -\alpha-2p$ ($D=4$) and $\alpha \rightarrow -\alpha -4p$ ($D=3$), for any $p$-form.
  \label{table2}}
\end{table}

The ten-dimensional origin of the $D$-dimensional $p$-form fields given in Table \ref{table2} resides not only in 10D $p$-forms but also in 10D mixed-symmetry fields. These mixed symmetry fields  must satisfy the following restricted reduction rule
in order to give rise to $p$-forms corresponding to half-supersymmetric branes
\cite{Bergshoeff:2011se}:\,\footnote{This rule is essentially a translation, in terms of indices, of the fact that supersymmetric branes
occur in representations whose highest weight is a real root.}
\vskip .4truecm

\noindent {\bf restricted reduction rule}\,:\ \ \ Consider a mixed-symmetry field $A_{m, n}$, with $m>n$,  corresponding to  a two-column Young tableaux with $m$ and $n$ boxes, respectively\,\footnote{The rule has an obvious extension to include fields $A_{n,n_1 ,n_2 ,...}$ with a mixed-symmetry structure corresponding to  multi-column Young tableaux. The $n_p$ indices have to be internal and parallel to $n_p$ of the $n_{p-1}$ indices.
When we write the mixed symmetry field $A_{n,n_1 ,n_2,...}$ we always assume that $n \geq n_1 \geq n_2 ...$, otherwise the field simply does not exist.}.
Upon toroidal reduction, this mixed-symmetry field gives rise to a potential corresponding to a
half-supersymmetric brane, provided that
 the $n$ indices are internal and along directions parallel to
$n$ of the $m$ indices.
\vskip .1truecm

We now review  the ten-dimensional IIA and IIB  fields, both form fields and mixed-symmetry fields, that give rise, after the restricted dimensional reduction
 rule stated above,  to the branes associated to the fields given in Table \ref{table2} using the light-cone rule. For the $\alpha=0$ fields we have
 \begin{equation}
\alpha=0: \qquad \quad B_2 \qquad {\rm metric}
\quad ,
\end{equation}
for both IIA and IIB, while for the $\alpha=-1$ fields we have
 \begin{equation}
{\rm IIA}: \qquad C_{2n+1} \qquad \qquad {\rm IIB}: \qquad C_{2n} \quad .
\end{equation}
Similarly, we know that the $\alpha=-2$ fields come from \cite{Bergshoeff:2011mh}

\begin{equation}
D_{6+n,n} \label{Dfieldsalphasquared2inIIAIIB}
\end{equation}
for both IIA and IIB.

We also know that the $\alpha=-3$ fields come from \cite{Bergshoeff:2011se}
  \begin{equation}
{\rm IIA}: \qquad E_{8+n, 2m+1,n} \qquad \qquad {\rm IIB}: \qquad E_{8+n, 2m,n}\quad .
\end{equation}
The resulting fields are listed in Tables \ref{EfieldsIIA} and \ref{EfieldsIIB}.

\begin{table}
\begin{center}
\begin{tabular}{|c||c|c|c|c|}
\hline \rule[-1mm]{0mm}{6mm} & $m=0$ & $m=1$ & $m=2$ & $m=3$\\
\hline
\hline \rule[-1mm]{0mm}{6mm}
$n=0$ & $E_{8,1}$ & $E_{8,3}$ & $E_{8,5}$ & $ E_{8,7}$ \\[.1truecm]
\hline
$n=1$ & $E_{9,1,1}$ & $E_{9,3,1}$ & $E_{9,5,1}$ &$E_{9,7,1}$ \\[.1truecm]
\hline \rule[-1mm]{0mm}{6mm}
$n=2$ & $-$ & $E_{10,3,2}$ & $E_{10,5,2}$ & $ E_{10,7,2}$ \\[.1truecm]
\hline
\end{tabular}
\end{center}
  \caption{\sl \footnotesize The $E_{8+n,2m+1,n}$ mixed symmetry fields of the IIA theory.
  \label{EfieldsIIA}}
\end{table}

\begin{table}
\begin{center}
\begin{tabular}{|c||c|c|c|c|}
\hline \rule[-1mm]{0mm}{6mm} & $m=0$ & $m=1$ & $m=2$ & $m=3$\\
\hline
\hline \rule[-1mm]{0mm}{6mm}
$n=0$ & $E_{8}$ & $E_{8,2}$ & $E_{8,4}$ & $ E_{8,6}$ \\[.1truecm]
\hline
$n=1$ & $-$ & $E_{9,2,1}$ & $E_{9,4,1}$ &$E_{9,6,1}$ \\[.1truecm]
\hline \rule[-1mm]{0mm}{6mm}
$n=2$ & $-$ & $E_{10,2,2}$ & $E_{10,4,2}$ & $ E_{10,6,2}$ \\[.1truecm]
\hline
\end{tabular}
\end{center}
  \caption{\sl \footnotesize The $E_{8+n,2m,n}$ mixed symmetry fields of the IIB theory.
  \label{EfieldsIIB}}
\end{table}

We next consider the $\alpha=-4$ fields. The fields in the last row above the double horizontal line of Table \ref{table2}, that is $F_{D, A_1 ...A_d}^+$, are generated by the following mixed-symmetry fields:
  \begin{equation}
  {\rm IIA}: \qquad F_{10, 2n+1 , 2n+1} \qquad \qquad {\rm IIB}: \qquad F_{10, 2n ,2n} \quad .
\end{equation}
Explicitly, these are the fields
 \begin{equation}
 F_{10,1,1} \qquad F_{10,3,3} \qquad F_{10,5,5} \qquad F_{10,7,7}
\end{equation}
in IIA and
 \begin{equation}
 F_{10} \qquad F_{10,2,2} \qquad F_{10,4,4} \qquad F_{10,6,6}
\end{equation}
in IIB. Observe that only in the IIB case the list contains a form, which is indeed the $\alpha=-4$ 9-brane that is the S-dual of the D9-brane.

We now consider the $F$ fields below the double horizontal line in Table \ref{table2}. These are the fields that do not have any 10D brane origin. It turns out that all these fields have a common IIA and IIB origin.
The fields $F_{D-1, A_1 ...A_{d-3}}$ and $F_{D,A, B_1 ...B_{d-3}}$  arise from the following series of mixed symmetry fields
\begin{equation}
F_{9+n, 3+m, m,n} \quad .
\end{equation}
The other $\alpha=-4$ fields $F_{D-2, A_1 ...A_{d-6}}$ and $F_{D-1, A, B_1 ...B_{d-6}}$ are generated by another series
of mixed-symmetry fields:
  \begin{equation}
F_{8+n, 6+m, m,n }
\quad .
\end{equation}
Explicitly, the resulting fields are given in Tables \ref{Ffieldssecondseries} and \ref{Ffieldsthirdseries}.

 \begin{table}
\begin{center}
\begin{tabular}{|c||c|c|c|c|c|}
\hline \rule[-1mm]{0mm}{6mm} & $m=0$ & $m=1$ & $m=2$ & $m=3$ & $m=4$\\
\hline
\hline \rule[-1mm]{0mm}{6mm}
$n=0$ & $F_{9,3}$ & $F_{9,4,1}$ & $F_{9,5,2}$ & $ F_{9,6,3}$ & $F_{9,7,4}$\\[.1truecm]
\hline
$n=1$ & $-$ & $F_{10,4,1,1}$ & $F_{10,5,2,1}$ &$F_{10,6,3,1}$ & $F_{0,7,4,1}$  \\[.1truecm]
\hline
\end{tabular}
\end{center}
  \caption{\sl \footnotesize The $F_{9+n,3+m,m,n}$ mixed symmetry fields of the IIA and IIB theory. They give rise, after restricted dimensional reduction, to the branes corresponding to the $D$-dimensional fields  $F_{D-1, A_1 ...A_{d-3}}$ and $F_{D,A, B_1 ...B_{d-3}}$.
  \label{Ffieldssecondseries}}
\end{table}

 \begin{table}
\begin{center}
\begin{tabular}{|c||c|c|}
\hline \rule[-1mm]{0mm}{6mm} & $m=0$ & $m=1$ \\
\hline
\hline \rule[-1mm]{0mm}{6mm}
$n=0$ & $F_{8,6}$  & $F_{8,7,1}$ \\[.1truecm]
\hline
$n=1$ & $-$ & $F_{9,7,1,1}$ \\[.1truecm]
\hline
\end{tabular}
\end{center}
  \caption{\sl \footnotesize The $F_{8+n,6+m,m,n}$ mixed symmetry fields of the IIA and IIB theory. They give rise, after restricted dimensional reduction, to the branes corresponding to the $D$-dimensional fields  $F_{D-2, A_1 ...A_{d-6}}$ and $F_{D-1, A, B_1 ...B_{d-6}}$ .
  \label{Ffieldsthirdseries}}
\end{table}

We  next consider the  $G$ fields with $\alpha=-5$ in Table \ref{table2}. In this case there are two series:
\begin{eqnarray}
& & {\rm series \ \ 1}: \qquad G_{D, A_1 ...A_{d-4} , \dot{a}}\,, \nonumber \\
& & {\rm series \ \ 2}: \qquad G_{D-1, A_1 ...A_{d-6} , a} \quad G_{D, A, B_1 ...B_{d-6}, a} \quad .
\end{eqnarray}
In the IIA theory, the first series is generated by the fields
 \begin{equation}
 G_{10, 4+n,2m+1, n}\quad ,
\end{equation}
while the second series is generated by
 \begin{equation}
G_{9+p, 6+n, 2m,n,p}\quad .
\end{equation}
Similarly, in the IIB theory the first series arises from
 \begin{equation}
 G_{10, 4+n,1+2m, n}\quad ,
\end{equation}
while the second series is generated by
 \begin{equation}
G_{9+p, 6+n, 2m+1,n,p}\quad .
\end{equation}
The explicit expressions for the fields are summarised in Tables \ref{GfieldsfirstseriesIIA}-\ref{GfieldssecondseriesIIB}.

 \begin{table}
\begin{center}
\begin{tabular}{|c||c|c|c|c|}
\hline \rule[-1mm]{0mm}{6mm} & $m=0$ & $m=1$ & $m=2$ & $m=3$ \\
\hline
\hline \rule[-1mm]{0mm}{6mm}
$n=0$ & $G_{10,4,1}$ & $G_{10,4,3}$ & $-$ & $-$\\[.1truecm]
\hline
$n=1$ & $G_{10,5,1,1}$ & $G_{10,5,3,1}$ & $G_{10,5,5,1}$ &$-$ \\[.1truecm]
\hline
$n=2$ & $-$ & $G_{10,6,3,2}$ & $G_{10,6,5,2}$ & $-$ \\[.1truecm]
\hline
$n=3$ & $-$ & $G_{10,7,3,3}$ & $G_{10,7,5,3}$ & $G_{10,7,7,3}$ \\[.1truecm]
\hline
\end{tabular}
\end{center}
  \caption{\sl \footnotesize The $ G_{10, 4+n,2m+1, n}$  mixed symmetry fields of the IIA theory. They give rise, after restricted dimensional reduction, to the branes corresponding to the $D$-dimensional fields $ G_{D, A_1 ...A_{d-4} , \dot{a}}$.
  \label{GfieldsfirstseriesIIA}}
\end{table}

 \begin{table}
\begin{center}
\begin{tabular}{|c||c|c|c|c|}
\hline \rule[-1mm]{0mm}{6mm} & $m=0$ & $m=1$ & $m=2$ & $m=3$ \\
\hline
\hline \rule[-1mm]{0mm}{6mm}
$p=0 \ \ n=0$ & $G_{9,6}$ & $G_{9,6,2}$ & $G_{9,6,4}$ & $G_{9,6,6}$\\[.1truecm]
\hline
$p=0 \ \  n=1$ & $-$ & $G_{9,7,2,1}$ & $G_{9,7,4,1}$ &$G_{9,7,6,1}$ \\[.1truecm]
\hline
$p=1 \ \ n=1$ & $-$ & $G_{10,7,2,1,1}$ & $G_{10,7,4,1,1}$ & $G_{10,7,6,1,1}$ \\[.1truecm]
\hline
\end{tabular}
\end{center}
  \caption{\sl \footnotesize The $ G_{9+p, 6+n,2m, n,p}$  mixed symmetry fields of the IIA theory. They give rise, after restricted dimensional reduction, to the branes corresponding to the $D$-dimensional fields $ G_{D-1, A_1 ...A_{d-6} , a} $ and $G_{D, A, B_1 ...B_{d-6}, a}$.
  \label{GfieldssecondseriesIIA}}
\end{table}

 \begin{table}
\begin{center}
\begin{tabular}{|c||c|c|c|c|}
\hline \rule[-1mm]{0mm}{6mm} & $m=0$ & $m=1$ & $m=2$ & $m=3$ \\
\hline
\hline \rule[-1mm]{0mm}{6mm}
$n=0$ & $G_{10,4}$ & $G_{10,4,2}$ & $G_{10,4,4}$ & $-$\\[.1truecm]
\hline
$n=1$ & $-$ & $G_{10,5,2,1}$ & $G_{10,5,4,1}$ &$-$ \\[.1truecm]
\hline
$n=2$ & $-$ & $G_{10,6,2,2}$ & $G_{10,6,4,2}$ & $G_{10,6,6,2}$ \\[.1truecm]
\hline
$n=3$ & $-$ & $-$ & $G_{10,7,4,3}$ & $G_{10,7,6,3}$ \\[.1truecm]
\hline
\end{tabular}
\end{center}
  \caption{\sl \footnotesize The $ G_{10, 4+n,2m, n}$  mixed symmetry fields of the IIB theory. They give rise, after restricted dimensional reduction, to the branes corresponding to the $D$-dimensional fields $ G_{D, A_1 ...A_{d-4} , \dot{a}}$.
  \label{GfieldsfirstseriesIIB}}
\end{table}

 \begin{table}
\begin{center}
\begin{tabular}{|c||c|c|c|c|}
\hline \rule[-1mm]{0mm}{6mm} & $m=0$ & $m=1$ & $m=2$ & $m=3$ \\
\hline
\hline \rule[-1mm]{0mm}{6mm}
$p=0 \ \ n=0$ & $G_{9,6,1}$ & $G_{9,6,3}$ & $G_{9,6,5}$ & $-$\\[.1truecm]
\hline
$p=0 \ \  n=1$ & $G_{9,7,1,1}$ & $G_{9,7,3,1}$ &$G_{9,7,5,1}$ & $G_{9,7,7,1}$ \\[.1truecm]
\hline
$p=1 \ \ n=1$ & $G_{10,7,1,1,1}$ & $G_{10,7,3,1,1}$ & $G_{10,7,5,1,1}$ & $G_{10,7,7,1,1}$ \\[.1truecm]
\hline
\end{tabular}
\end{center}
  \caption{\sl \footnotesize The $ G_{9+p, 6+n,2m+1, n,p}$  mixed symmetry fields of the IIB theory. They give rise, after restricted dimensional reduction, to the branes corresponding to the $D$-dimensional fields $ G_{D-1, A_1 ...A_{d-6} , a} $ and $G_{D, A, B_1 ...B_{d-6}, a}$.
  \label{GfieldssecondseriesIIB}}
\end{table}

Finally, we consider the $H$ fields in Table \ref{table2} with $\alpha=-6$. The fields $H_{4, A_1 ...A_4}$ and $H_{3, A, B_1 ...B_5}$ corresponding to the $\alpha=-6$ space-filling branes in 4D and 3D respectively, are generated both in IIA and in IIB by the mixed symmetry fields
\begin{equation}
H_{10, 6+n,2+m,m,n}\quad ,
\end{equation}
whose explicit expression is given in Table \ref{Hfieldsfirstseries}. The $\alpha=-6$ domain walls in 3 dimensions, associated to the field $H_{2,A_1 A_2 A_3}$ arise instead from the mixed symmetry fields
\begin{equation}
H_{9,7,4+n,n}
\end{equation}
in both the IIA and the IIB theory, once the restricted reduction rule is applied. One can derive in a similar way the fields associated to the branes with lower $\alpha$ in $D=4$ and $D=3$, but this is straightforward given that all these branes are related to the previous ones by S-duality.

 \begin{table}
\begin{center}
\begin{tabular}{|c||c|c|c|c|c|c|}
\hline \rule[-1mm]{0mm}{6mm} & $m=0$ & $m=1$ & $m=2$ & $m=3$ & $m=4$ & $m=5$\\
\hline
\hline \rule[-1mm]{0mm}{6mm}
$n=0$ & $H_{10,6,2}$ & $H_{10,6,3,1}$ & $H_{10,6,4,2}$ & $H_{10,6,5,3}$ & $H_{10,6,6,4}$ & $-$\\[.1truecm]
\hline
$n=1$ & $-$ & $H_{10,7,3,1,1}$ & $H_{10,7,4,2,1}$ &$H_{10,7,5,3,1}$ & $H_{10,7,6,4,1}$ & $H_{10,7,7,5,1}$ \\[.1truecm]
\hline
\end{tabular}
\end{center}
  \caption{\sl \footnotesize The $ H_{10, 6+n,2+m, m,n}$  mixed symmetry fields of the IIA and  IIB theories. They give rise, after restricted dimensional reduction, to the branes corresponding to the $\alpha=-6$  fields $H_{4, A_1 ...A_4} $ in 4D and $H_{3, A, B_1 ...B_5}$ in 3D.
  \label{Hfieldsfirstseries}}
\end{table}

This finishes our review on the
enumeration of all half-supersymmetric branes of $D$-dimensional maximal supergravity together with their 10D origin in terms of forms and mixed-symmetry fields.

\subsection{Heterotic truncation}

We are now in a position to discuss  the heterotic truncation of the  IIA/IIB fields discussed in the previous subsection to those of  the
ten-dimensional $\text{SO}(8,8)^{+++}$  theory.  We first discuss the truncation at the level of branes or real roots. Next, we discuss its
extension to the full spectrum.

\subsubsection{Brane truncation}

The truncation at the level of the real roots
is very simple and reads
\bigskip

\noindent {\bf heterotic truncation rule:}\ \ truncate all fields of the IIA/IIB theory that are part of IIA or IIB but not both. Equivalently, keep only the real roots of $\text{E}_8^{+++}$ that are common to the IIA and the IIB theory.
\bigskip

After toroidal compactification, the heterotic truncation
rule  implies
that we truncate all fields in Table \ref{table2} that have either a IIA or IIB origin but not both.
The truncation projects out all the odd $\alpha$ branes, but also the $\alpha=-4$  branes corresponding to  $F_{D, A_1 ...A_d}^+ $.  Note that these  are the only even $\alpha$ branes that have a {\sl different} IIA and IIB origin. This explains why these branes are not present in the half maximal theory, as discussed in the previous section.
 Remarkably, this simple prescription automatically eliminates all branes whose worldvolume dimension is higher than 6. This is consistent with the
 fact that there are no matter multiplets with 8 supercharges in spacetime dimensions higher than 6.

\subsubsection{Spectrum truncation}

One may wish to extend the above heterotic truncation rule to include not only the fields corresponding to the supersymmetric branes, i.e.~the
real roots of $\text{E}_8^{+++}$,   but
to include the {\sl full} Kac-Moody spectrum, including the fields corresponding to null and imaginary roots, thereby truncating the spectrum of
$\text{E}_8^{+++}$ to that of $\text{SO}(8,8)^{+++}$. In \cite{Kleinschmidt:2004dy} it was shown that the  $\text{SO}(8,8)^{+++}$ is contained in the even-$\alpha$ spectrum of   $\text{E}_8^{+++}$\,\footnote{The authors actually considered the over-extended algebras  $\text{SO}(8,8)^{++}$ and   $\text{E}_8^{++}$, but the analysis is similar.}.  In the previous subsection we have managed to further characterize this truncation as far  as the branes are concerned, but as we will see
it turns out that this  is not so easy when the whole spectrum is taken into account.

As an example, we consider all the 10-dimensional $\text{SO}(8,8)^{+++}$ fields (forms and mixed-symmetry fields; real, null and imaginary roots) that give rise to forms in
$D\ge 6$ dimensions. These fields are  (we put in brackets the squared length of the corresponding root $\alpha^2$)\,~\footnote{This should not be confused with the dilaton weight $\alpha$.}
\begin{eqnarray}
& & B_2 \ \ (2) \quad D_6 \ \ (2)  \quad  D_8 \ \ (0)  \quad  D_{7,1} \ \ (2)  \quad  D_{10} \ \ (-2)  \quad   D_{9,1} \ \ (0)  \quad   D_{8,2} \ \ (2)  \nonumber \\
& &    D_{10,2} \ \ (0)  \quad
 D_{9,3} \ \ (2)  \quad  D_{10,4} \ \ (2) \quad   F_{9,3} \ \ (2)  \quad  2 \times  F_{10,4} \ \ (-2)  \quad   F_{10,3,1} \ \ (0)   \nonumber \\
& &    F_{9,4,1} \ \ (2)  \quad   F_{10,4,2} \ \ (0)
\quad  F_{10,4,1,1} \ \ (2)  \quad   . \label{allhetfieldsD=10}
\end{eqnarray}
The $\alpha^2=2$ fields above are those discussed previously. We now wish  to obtain {\sl all} fields given above from a truncation of the $\text{E}_{8}^{+++}$ fields of IIA and IIB. First of all, we have to project on the fields with even $\alpha$, that is the $B$, $D$ and $F$ fields. If we do this, we realise that the IIA fields that are left and that do not belong to eq.~\eqref{allhetfieldsD=10} are\,\footnote{We have given the $D_{10}$ field a prime to distinguish it from the other $D_{10}$ field that
already occurs in eq.~\eqref{allhetfieldsD=10}. Indeed in both the IIA and IIB theory there are two such fields, while there is only one in eq.~\eqref{allhetfieldsD=10}. }
  \begin{equation}
 D^\prime_{10} \  \ (-2) \quad F_{10,1,1} \ \ (2) \quad F_{10,3,1} \ \ (0) \quad F_{10,3,3} \ \ (2) \quad , \label{projectedoutallfieldsIIA}
\end{equation}
while for the IIB fields we find
  \begin{equation}
 D^\prime_{10} \  \ (-2) \quad F_{10} \ \ (2) \quad F_{10,2} \ \ (0) \quad F_{10,4} \ \ (-2) \quad F_{10,2,2} \ \ (2) \quad  F_{10,4,2} \ \ (0) \quad F_{10,4,4} \ \ (2) \quad .\label{projectedoutallfieldsIIB}
\end{equation}
The $F$-fields generalise the same pattern that we have already seen for the $\alpha^2=2$ fields, i.e.~the ones corresponding to the real roots, discussed above. They are even $\alpha$ fields that nevertheless are {\sl not}
common to IIA and IIB and, therefore, must be truncated: on top of projecting out the odd $\alpha$ fields, we also truncate the even $\alpha$  states fields that are not contained in the intersection between IIA and IIB.
However, applying the heterotic truncation rule stated above is not enough: the $D_{10}^\prime$ field is common to IIA and IIB but, nevertheless must be truncated, since it does not occur in eq.~\eqref{allhetfieldsD=10}.

One can extend the above analysis and consider also the fields in 10 dimensions that give rise to forms in 5, 4 and 3 dimensions. In this case the situation is even  more complicated and there seems to be no pattern at all. We find that the intersection rule (common IIA/IIB origin) is violated by more fields. Furthermore, there are  now also even $\alpha$  $(D-1)$ forms that must be projected out, while one can see that the
even $\alpha$ fields in  eqs.~\eqref{projectedoutallfieldsIIA} and \eqref{projectedoutallfieldsIIB} only contribute to the $D$-forms.

\subsection{Gravity and matter sector}

In this subsection we justify the assumption made in the previous analysis that one can consider the 10-dimensional spectrum of the $\text{SO}(8,8)^{+++}$ Kac-Moody algebra as the gravity sector of the $\text{SO}(8,8+n)^{+++}$ Kac-Moody algebra. In order to do this, we will consider the spectrum of form fields (that is all fields with antisymmetric spacetime indices) that arises in the six-dimensional  ${\cal N} = (1,1)$  theory.  For any $n$, this spectrum is given  by~\footnote{Here we denote in brackets the squared length of the root associated to the highest weight of the $\text{SO}(4,4+n)$ representation.}
   \begin{eqnarray}
 & &  B_{1, A}  \  (2) \quad  B_2 \  (2) \quad   D_2 \  (2) \quad  D_{3,A}  \ (2) \quad   D_4  \ (0) \quad  D_{4, A_1 A_2}  \ (2) \quad    D_{5, A}  \ (0) \nonumber \\
& &     D_{5, A_1 A_2 A_3 }  \ (2) \quad   D_{6}  \ (-2) \quad   D_{6, A_1 A_2}  \ (0) \quad   D_{6, A_1 A_2 A_3 A_4}  \ (2) \quad  F_{5,A} \ (2) \nonumber\\
& & 2 \times F_{6} \ (-2) \quad F_{6, A_1 A_2} \ (0) \quad F_{6 ,AB} \ (2)\,. \label{sixdimhetforms}
\end{eqnarray}
 Upon six-dimensional level decomposition the $\text{SO}(8,8)^{+++}$ Kac-Moody algebra gives this spectrum for $n=0$, i.e.~all fields are in representations of the symmetry $\text{SO}(4,4)$. The  $\text{SO}(8,8+n)^{+++}$ Kac-Moody algebra gives the same spectrum, but this time with
 all fields in representations of  $\text{SO}(4,4+n)$.  Therefore, the full Kac-Moody spectrum generalises what the standard supergravity spectrum does: the dimensional reduction of the pure supergravity theory in 10D and the dimensional reduction of the supergravity theory plus vector multiplets give the same $D$-dimensional theory but with different amounts of $D$-dimensional vector multiplets.

When $n=0$, the fields in eq.~\eqref{sixdimhetforms}
 arise from the dimensional reduction of the 10-dimensional mixed-symmetry fields given in eq.~\eqref{allhetfieldsD=10}. We will restrict for simplicity our attention to the $\alpha=0$ field $B_2$ and the $\alpha=-2$ fields. The latter can be written in the compact notation
  \begin{equation}
 D_{6+n+2m,n } \ \ (\alpha^2 = 2 -2m)\quad ,\label{alpha0alpha-288+++}
\end{equation}
as can be seen from  eq.~\eqref{allhetfieldsD=10}.
In particular the $D$ fields associated with the real roots are those  with $m=0$, that are indeed the fields in
eq.~\eqref{Dfieldsalphasquared2inIIAIIB}. The reader can see that reducing all these fields (and the graviton) on $T^4$ gives the six-dimensional $\alpha=0$ and $\alpha=-2$ form  fields in eq.~\eqref{sixdimhetforms} as representations of $\text{SO}(4,4)$.
We can now  consider the ten-dimensional mixed-symmetry fields that are in the  $\text{SO}(8,8+n)^{+++}$ Kac-Moody algebra but not in the  $\text{SO}(8,8)^{+++}$ Kac-Moody algebra. These fields are basically the Kac-Moody generalisation of the bosonic sector of the ten-dimensional vector multiplet. We first consider the $\alpha=0$ fields.
The ten-dimensional $\text{SO}(8,8+n)^{+++}$ algebra gives an $\text{SO}(n)$ symmetry, and the
 $\alpha=0$ fields that are in $\text{SO}(8,8+n)^{+++}$ algebra and not in eq.~\eqref{alpha0alpha-288+++} are  clearly the 1-forms  $B_{1,A}$, where the index $A$ is the vector index of $\text{SO}(n)$. The fields $B_{1,A}$ thus belong to the matter sector.
It is straightforward to verify that the dimensional reduction of these matter  fields, together with $B_2$ and the graviton, gives $8+n$ 1-forms that form a vector of $\text{SO}(4,4+n)$. This is indeed the standard supergravity result.

We next consider the $\alpha=-2$ fields.  The SO$(n)$ representations of all the  $D$ fields are given by the compact formula
  \begin{equation}
  D_{6+n+2m +p, n, A_1 ...A_p} \  \ \ p=1,2,3,4 \quad .
\end{equation}
This includes the $n=0$ fields given in \eqref{alpha0alpha-288+++}. The other fields, with $n\ne0$, all belong to the matter sector.
The reader can check that these 10-dimensional fields, as representations of $\text{SO}(n)$, when dimensionally reduced on $T^4$ promote the form fields with $\alpha=-2$ in eq.~\eqref{sixdimhetforms} from representations of $\text{SO}(4,4)$ to representations of $\text{SO}(4,4+n)$. A similar analysis can be done for the fields with lower $\alpha$, but the analysis is more complicated.

\section{Duality and K3 Wrapping Rules}

The IIA theory compactified on K3 and the heterotic theory on $T^4$
are  conjectured to be S-dual \cite{Hull:1994ys}. In this section we want to consider this duality from the point of view of the corresponding Kac-Moody algebras.
We will consider the orbifold limit $T^4/{\mathbb Z}_2$ of K3. In this limit one can consider a truncation of the low-energy action of the IIA theory in which one compactifies over $T^4$ keeping only the fields that wrap over even cycles. This gives ${\cal N}=(1,1)$ supergravity in six dimensions coupled to four vector multiplets, which is the low-energy limit of the untwisted sector of the IIA theory compactified on the orbifold. Here we want to first extend this result at the level of the Kac-Moody algebra, which corresponds to the inclusion of form fields of all rank in the six-dimensional theory. We thus consider the ten-dimensional IIA $\text{E}_8^{+++}$ mixed-symmetry fields and we compactify them on $T^4$ keeping only the fields that wrap on even cycles. We keep only the fields that give rise to forms in six dimensions, and we compare the result with the spectrum of six-dimensional forms resulting from $\text{SO}(8,8)^{+++}$, corresponding to the reduction of the gravity sector of the heterotic spectrum in ten dimensions. We then restrict our attention to the branes, and show how the wrapping rules that the IIA and IIB branes satisfy when compactified on the torus are generalised to wrapping rules on the orbifold.

\subsection{IIA string theory on K3}

We consider the particular orbifold limit in which K3 is $T^4 /
\mathbb{Z}_2$. In this limit, the untwisted sector of the IIA theory
gives the ${\cal N} = (1,1)$ supergravity multiplet plus four vector
multiplets. The orbifold has 16 fixed points, corresponding to a twisted sector giving rise to 16 additional vector multiplets.
Here we want to consider all the forms in the untwisted
sector (real, null and imaginary roots) as obtained from a particular reduction of the IIA $\text{E}_{8}^{+++}$
fields (forms as well as mixed-symmetry fields)  in 10 dimensions. In the previous section we
have shown that the 10-dimensional fields of the Kac-Moody algebra
$\text{SO}(8,8)^{+++}$ that upon reduction on $T^4$  give rise to the forms in six dimensions given in eq.  \eqref{sixdimhetforms}, are those in
eq.~\eqref{allhetfieldsD=10}.
The 6D forms are written as representations of $\text{SO}(4,4)$, which is  the symmetry of the non-chiral six-dimensional $\text{SO}(8,8)^{+++}$ theory. On the other hand, the same fields, albeit as representations of $\text{SO}(4,20)$, occur if one considers the six-dimensional theory arising from  $\text{SO}(8,24)^{+++}$. We now conjecture that the untwisted sector of the IIA $\text{E}_8^{+++}$  theory on the orbifold should give the fields that are S-dual to those in eq.~\eqref{sixdimhetforms} as representations of $\text{SO}(4,4)$, while the twisted sector will extend these representations to representations of $\text{SO}(4,20)$, exactly as it happens in the S-dual heterotic theory as discussed in Subsection 3.3.

To check the conjecture we must compute the forms arising from reducing the mixed-symmetry fields of IIA on $T^4/\mathbb{Z}_2$.
 The orbifold projection is taken as follows: we first perform  a standard dimensional reduction of IIA on  $T^4$ and, next, we take only the fields with an {\sl even} number of the internal $\text{GL}(4,\mathbb{R})$ indices. This is the same as saying that we only reduce over even cycles, i.e.~2-cycles and 4-cycles. Roughly speaking, ``taking even cycles'' is dual to ``taking even values of $\alpha$''. More precisely, the duality is  equivalent to the statement that first performing a heterotic truncation of IIA and next reducing over $T^4$ is equivalent to first reducing IIA  over $T^4$ and next performing the orbifold projection.
 Note that the heterotic truncation and the orbifold projection  are rather different in nature, and therefore, the duality is non-trivial.

 Given the orbifold projection, our task is now to show that, after reduction over  $T^4/\mathbb{Z}_2$, for each form the corresponding $\text{GL}(4,\mathbb{R})$ representations add up to representations of $\text{SO}(4,4)$. We start by considering the scalars. These can only arise from the metric (10 of them) and from the 2-form (6 of them) for a total of 16 scalars parametrising $\text{SO}(4,4) / \text{SO}(4) \times \text{SO}(4)$. The additional scalar is the ten-dimensional dilaton. We next consider each form separately.
\bigskip

\noindent  {\it 1-forms}\ \ \  The 1-forms can only come from the RR
IIA fields $A_1$, $A_3$ and $A_5$. This gives the
$\text{SL}(4,\mathbb{R})$ representations
\begin{equation}
{\bf 1} \oplus {\bf 6} \oplus {\bf 1} \quad .
\end{equation}
In decomposing the $\text{SO}(4,4)$ representations in terms of $\text{SL}(4,\mathbb{R}) \subset \text{SO}(4,4)$, one has
\begin{equation}
{\bf 8} = {\bf 1} \oplus {\bf 6} \oplus {\bf 1} \quad ,
\end{equation}
which means that the 1-forms give a field $A_{1,A}$ in agreement with eq.~\eqref{sixdimhetforms}.
\bigskip

\noindent {\it 2-forms}\ \ \  The 2-forms come from $A_2$ and $A_6$,
and give two singlets, in agreement with  eq.~\eqref{sixdimhetforms}.
\bigskip

\noindent {\it 3-forms}\ \ \  The 3-forms come from $A_3$, $A_5$ and
$A_7$ and the computation is identical to the 1-forms, of which they
are the dual.
\bigskip

\noindent {\it 4-forms}\ \ \  The fields that give 4-forms are $A_6$
$({\bf 6})$, $A_8$ $({\bf 1})$,  $A_{7,1} $ $({\bf 15 \oplus 1})$
and $A_{8,2}$ $({\bf 6})$. We have
\begin{equation}
{\bf 28} = {\bf 1} \oplus {\bf 6} \oplus {\bf 6} \oplus {\bf 15} \quad ,
\end{equation}
where the ${\bf 28}$ is the adjoint (two antisymmetric indices) of $\text{SO}(4,4)$, which means that the 4-forms are
 \begin{equation}
A_4 \quad \quad A_{4, A_1 A_2}\,.
\end{equation}
This is again in agreement with  eq.~\eqref{sixdimhetforms}.
\bigskip

\noindent {\it 5-forms}\ \ \  From  eq.~\eqref{sixdimhetforms} we expect
the fields to collect in the ${\bf 8 \oplus 8 \oplus 56}$, where the
${\bf 56}$ corresponds to 3 antisymmetric indices which decomposes
as
  \begin{equation}
{\bf 56} = {\bf 15 \oplus 15 \oplus 6 \oplus 10 \oplus \overline{10}}
\end{equation}
under  $\text{SL}(4,\mathbb{R}) \subset \text{SO}(4,4)$.
The IIA fields are $A_5$  $({\bf 1})$,  $A_7$  $({\bf 6})$,  $A_9$  $({\bf 1})$,  $A_{8,1}$  $({\bf 15 \oplus 1})$,  $A_{9,2}$  $({\bf 6})$,  $A_{9,3,1}$  $({\bf 15})$, $A_{8,3}$  $({\bf 6 \oplus \overline{10}})$,  $A_{9,4}$  $({\bf 1})$ and $A_{9,1,1}$  $({\bf 10})$.  It is easy to check that these fields give indeed the correct $\text{SO}(4,4)$ representations.
\bigskip

\noindent {\it 6-forms}\ \ \ As we will see, for the 6-forms one has
to project out some extra fields because reducing on $T^4$ and taking only
the fields with an even number of internal indices gives too many
fields. Remarkably, we will see that the fields that one has to
project out are precisely $A_{10}$, $A_{10,1,1}$, $A_{10,3,1}$ and
$A_{10,3,3}$ which we have already seen in the previous section as
the extra even $\alpha$ fields that had to be projected out in the heterotic truncation.

From  \eqref{sixdimhetforms} we expect the six-dimensional forms to collect in the representations
\begin{equation}
{\bf 1 \oplus 1 \oplus 1 \oplus 28 \oplus 28 \oplus 35_{\rm V} \oplus 35_{\rm S} \oplus 35_{\rm C}} \ \ . \label{sixformrepresentations}
\end{equation}
The ${\bf 35}$'s decompose under $\text{SL}(4,\mathbb{R})$ as
\begin{eqnarray}
& & {\bf 35}_{\rm V} = {\bf 1 \oplus 1 \oplus 1 \oplus 6 \oplus 6 \oplus 20^\prime} \nonumber \\
& & {\bf 35}_{\rm S,C}= {\bf 10 \oplus \overline{10}  \oplus 15 } \quad .
\end{eqnarray}
We now perform the reduction of the IIA fields. The 6-forms come from $A_6$ $({\bf 1})$,  $A_8$ $({\bf 6})$,   $A_{7,1}$ $({\bf 6 \oplus 10})$,  $A_{9,1}$ $({\bf 15\oplus 1})$,   $A_{8,2}$ $({\bf 20^\prime \oplus 15 \oplus 1})$,  $2 \times A_{10}$ $({\bf 1\oplus 1 })$,   $2 \times A_{9,3} $ $({\bf \overline{10} \oplus \overline{10} \oplus 6 \oplus 6})$,  $A_{10,2}$ $({\bf 6})$,  $2 \times A_{10,4}$ $({\bf 1 \oplus 1 })$,   $2 \times A_{10,3,1} $ $({\bf 15 \oplus 15})$,   $A_{9,4,1}$ $({\bf 15 \oplus 1 })$,  $A_{10,4} $ $({\bf 1})$,  $A_{10,4,2}$ $({\bf 6})$,  $A_{10,4,1,1}$ $({\bf 10})$,     $A_{10,1,1}$ $({\bf 10})$ and  $A_{10,3,3}$ $({\bf \overline{10}})$.

As anticipated, we recover the representations of eq.~\eqref{sixformrepresentations} provided that one of the 10-forms $A_{10}$, one of the two $A_{10,3,1}$ fields, and the $A_{10,1,1}$ and $A_{10,3,3}$ fields are projected out. These are exactly the fields in eq.~\eqref{projectedoutallfieldsIIA} that were projected out in the previous section as the only even $\alpha$ fields that do not survive the projection to the ten dimensional heterotic theory.

\subsection{K3 wrapping rules}

In the previous subsection we have seen that  the ``even $\alpha$ rule'' and the ``even cycle'' rule are not enough to establish the duality between the heterotic theory on $T^4$ and the IIA theory over K3. Both rules have to be supplemented with the truncation of an additional set of fields. Remarkably, we find that in both cases this set of additional fields is precisely the same. The Kac-Moody interpretation of these extra truncations is not yet clear to us.

Instead of discussing the duality between IIA on K3 and the heterotic theory on $T^4$ from the full Kac-Moody point of view, we may also consider the same duality at the level of the half-supersymmetric branes only. In doing this we will also consider the IIB theory on  K3, whose low-energy limit is the ${\cal N} =(2,0)$ chiral six-dimensional supergravity theory coupled to 21 tensor multiplets. Again, we will consider the orbifold limit $T^4/\mathbb{Z}_2$ of K3, in which case the untwisted sector gives supergravity plus 5  tensor multiplets, while the 16 additional tensor multiplets in the twisted sector are associated to the 16 fixed points. The branes of this theory were discussed at the end of subsection 2.1.

We are interested in those fields that, as representations of the global symmetry group, have a highest weight  that corresponds to real roots of the Kac-Moody algebra. As in the previous subsection, we do not specify the dilaton weight $\alpha$ (which is  not even defined for the (2,0) case), and generically denote all fields
with $A$. The fields of the $(1,1)$ theory (see Table \ref{halfsusybranefields}) are
  \begin{equation}
A_{1, A} \quad 2 \times A_2  \quad A_{3,A} \quad A_{4, A_1 A_2} \quad A_{5, A_1 A_2 A_3} \quad A_{5, A} \quad A_{6, A_1 ...A_4} \quad A_{6,AB} \quad ,
\end{equation}
while the fields of the $(2,0)$ theory are given in eq.~\eqref{fieldsinthe20theory}.
The half-supersymmetric branes follow from the fields using the light-cone rule.

 We first restrict our attention to the branes with worldvolume at most 4, i.e.~we do not consider domain walls and space-filling branes. The reason is that we want to derive all branes we consider from wrapping rules starting from the 10-dimensional branes, and we know that in the maximally supersymmetric case (that is IIA/IIB compactified on a torus)  starting from 7 dimensions there are domain walls  that do not arise from wrapping the 10-dimensional branes (see the review in subsection 3.1). In the $(1,1)$ theory  there are 8 lightlike directions, while in the $(2,0)$ theory the number of lightlike directions is 10. Using the representations of the fields given above and applying the light-cone rule we find the following number of (fundamental and solitonic) p-branes $(p \le 3)$
  \begin{eqnarray}
& &  {\rm 0-branes}: \  \    8 \nonumber \\
& & {\rm 1-branes}: \ \ 1+1 \nonumber \\
& & {\rm 2-branes}: \ \ 8 \nonumber \\
& & {\rm 3-branes}: \ \ 24  \label{branes11}
\end{eqnarray}
for the $(1,1)$ theory (see Tables \ref{Fwrapping} and \ref{Swrapping}) and
  \begin{eqnarray}
& & {\rm 1-branes}: \ \ 10 \nonumber \\
& & {\rm 3-branes}: \ \ 40   \label{branes20}
\end{eqnarray}
for the $(2,0)$ theory, see eq.~\eqref{branesiib}.

We now wish to verify whether the above spectrum of branes of the two theories in eqs.~\eqref{branes11} and \eqref{branes20} can be obtained
from the IIA and IIB theory by a set of K3 wrapping rules. We take these K3 wrapping rules to be the same wrapping rules as we used for the torus reduction.
For fundamental and solitonic branes these wrapping rules were given in
 eqs.~\eqref{Fbranewrapping} and \eqref{Sbranewrapping}. We will also need the D-brane ($\alpha=-1$) and E-brane ($\alpha=-3$) wrapping rules which are given by\,\footnote{The IIB theory also has a $\alpha=-4$ brane, the S-dual of the D9-brane. This a space-filling brane which can only wrap.
 Upon wrapping it doubles.  }
     \begin{eqnarray}\label{Dbranewrapping}
 & & {\rm D} \ \ \ \left\{ \begin{array}{l}
{\rm wrapped} \ \ \ \ \rightarrow\  \ \ {\rm undoubled}\\
{\rm unwrapped} \ \ \rightarrow \ \ {\rm undoubled}
 \end{array} \right.
 \end{eqnarray}
and
 \begin{eqnarray}\label{Ebranewrapping}
 & & {\rm E} \ \ \ \left\{ \begin{array}{l}
{\rm wrapped} \ \ \ \ \rightarrow\  \ \ {\rm doubled}\\
{\rm unwrapped} \ \ \rightarrow \ \ {\rm doubled} \quad ,
 \end{array} \right.
\end{eqnarray}
respectively. The only difference with the torus reduction is that we now  use the fact that the
 K3 manifold has no non-trivial 1-cycles, while there are non-trivial 2-cycles. This means that each 10-dimensional brane can be unwrapped, wrapped on a 2-cycle or wrapped on the whole K3 manifold.
 Furthermore, we assume that the branes see an ``effective'' number of 2-cycles $n$. It turns out that  this effective number of 2-cycles is the same in the IIA and IIB theories, which is $n=6$. In fact, this is just the number of ways that a 2-torus $T^2$ can be embedded in $T^4$, where we are considering the orbifold limit $T^4 / \mathbb{Z}_2$  of K3, and $\mathbb{Z}_2$ removes the odd cycles. The fact that in the IIA case the K3 wrapping rules, defined in this way, precisely reproduce the spectrum of branes given in eq.~\eqref{branes11}, as we will show below,  tells us that all half-supersymmetric  branes come from the untwisted sector. The twisted sector only gives additional compact directions in the global symmetry group which, according to the light-cone rule, does not lead to additional half-supersymmetric branes.

We first consider wrapping the branes of the IIA theory on K3. The 0-branes can only come from the D0 unwrapped, the D2 wrapped on a 2-cycle and the D4 wrapped on K3. Given that the D-branes never double, we get
 \begin{equation}
{\rm number \ of \ 0-branes} = 2+ n  \quad ,
\end{equation}
which comparing with the first line of eq.~\eqref{branes11} gives $n=6$.  We then consider the 1-branes. These can only come from the unwrapped fundamental string F1 and the NS5 wrapped on K3. The unwrapped fundamental string does not double, and the same is true for solitonic objects that are fully wrapped. This leads to
  \begin{equation}
{\rm number \ of \ 1-branes} = 1+1   \quad ,
\end{equation}
as in the second line of eq.~\eqref{branes11}. The 2-branes come from the D2 unwrapped, the D4 wrapped on a 2-cycle and the D6 on the whole K3 manifold. Again, this gives
 \begin{equation}
{\rm number \ of \ 2-branes} = 2+ n  \quad ,
\end{equation}
which again gives $n=6$. Finally, the 3-branes can only come from the NS5 wrapped on a 2-cycle. Since 2 directions of this solitonic brane are unwrapped, we get a factor 4.
This gives
 \begin{equation}
{\rm number \ of \ 3-branes} = 4  n  \quad ,
\end{equation}
and again this gives $n=6$ for consistency with the last line of eq.~\eqref{branes11}.

We now consider wrapping the branes of the IIB theory on K3. Since there are no odd cycles on K3, one cannot obtain $p$-branes with $p$ even (including $p=0$)  in the 6-dimensional theory. The 1-branes come from the F1 and the D1 unwrapped, the D3 wrapped on a 2-cycle and the D5 and the NS5 on the whole K3. No doubling is involved for these branes, because the F1 does not wrap and the solitonic NS5 is fully wrapped, while in general no doubling is involved for the D-branes. This leads to
  \begin{equation}
{\rm number \ of \ 1-branes} = 1 + 1 + n + 1 + 1    \quad ,
\end{equation}
which gives again $n=6$  by comparing with the first line of  eq.~\eqref{branes20}. Finally, we consider the 3-branes. These come from the D3 unwrapped, the D5 and NS5 on a 2-cycle, and finally the D7 and its S-dual E7 on the whole K3 manifold. The D-branes do not double, so one gets 1 from the D3, $n$ from the D5 and 1 from the D7, while the NS5 on the 2-cycles gives $4n$  branes (the factor 4 from the two unwrapped directions). Finally, the fully wrapped E7 doubles, but we have to remember that this counts as $1/2$ in the 10-dimensional theory because there is no such brane in the IIA theory. So one gets $1/2 \cdot 2^4 =8$. The final result is
  \begin{equation}
{\rm number \ of \ 3-branes} = 1+ n + 4n + 1 + 8  \quad ,
\end{equation}
and remarkably if $n=6$ we get 40 branes exactly as in the last line of eq.~\eqref{branes20}.

We have seen that the spectrum of 6D $p$-branes, with $0\le p \le 3$, is
precisely reproduced  by using  the K3 wrapping rules defined above.
We now consider also the 4-branes  and 5-branes.  In principle, we
do not expect all these branes to arise from ten dimensions using
our K3 wrapping rules, because we know already from the torus
dimensional reduction that starting from $D=7$ there are domain
walls that cannot be obtained from 10 dimensions. Nonetheless, the
consistency check is that using the K3 wrapping rules we should not get
{\sl more} branes than we have in six dimensions. The relevant fields in the $(1,1)$ theory
are given by the 5-form fields $A_{5, A_1 A_2 A_3}$ and $A_{5, A}$ and by
the 6-form fields $A_{6, A_1 ...A_4}$ and $A_{6, AB}$. Applying the light-cone rule these fields lead to the following number of branes (see Tables \ref{Swrapping} and \ref{Ffieldstable}):
  \begin{eqnarray}
& &  {\rm 4-branes}: \  \    32 +8\,, \nonumber \\
& & {\rm 5-branes}: \ \ 16+8\,.   \label{branes11extra}
\end{eqnarray}
In the  $(2,0)$ theory  we only have the 6-form fields $A_{6, A_1 A_2, B}$ and they lead to the following number of branes (see
eq.~\eqref{branesiib}):
  \begin{equation}
 {\rm 5-branes}: \ \ 80 \quad . \label{branes20extra}
\end{equation}

We now consider which of these branes come from the IIA and IIB theory by dimensional reduction using the K3 wrapping rules. In the IIA case we have that the 4-branes come from the D4 unwrapped (one brane), the D6 wrapped on a 2-cycle  ($n$ branes) and the  D8 wrapped on the whole of K3 (one brane). In total one gets
   \begin{equation}
{\rm number \ of \ 4-branes} =1 + n + 1 \quad ,
\end{equation}
and for $n=6$ one gets 8 branes. Although this is not the total number of branes, it is remarkable that one obtains exactly the branes that come from one of the two irreducible representations of $SO(4,4+n)$, as shown in the first row of eq.~\eqref{branes11extra}. The other representation should follow from the reduction of ``generalized'' Kaluza-Klein monopoles, see e.g.~\cite{Bergshoeff:2011se}. Similarly, for the 5-branes one obtains 16 branes coming from the (four times unwrapped) NS5 brane. That is
   \begin{equation}
{\rm number \ of \ 5-branes} =16 \quad ,
\end{equation}
and again  one obtains the branes coming from one of the two irreducible representations, as shown in the second line of eq.~\eqref{branes11extra}.

In the IIB case we get 16 5-branes from the unwrapped NS5, one 5-brane from D5, $n$ 5-branes from the D7 wrapped on a 2-cycle, $8n$ 5-branes coming from the E7 wrapped on a 2-cycle, one 5-brane coming from the D9 wrapped on K3 and 8 5-branes coming from the S-dual of the D9.
The result is
     \begin{equation}
{\rm number \ of \ 5-branes} =16 + 1 + n + 8n + 1 + 8\quad ,
\end{equation}
and remarkably this gives 80 branes if $n=6$, in agreement with eq.~\eqref{branes20extra}. In this case the representation is irreducible, and therefore we must get all the branes from the K3 wrapping rules.

This finishes our discussion of the K3 wrapping rules.

\section{Heterotic Branes and Central Charges}

In this section we study the relation between the number of half-supersymmetric heterotic branes and the central charges of the half-maximal
supersymmetry algebra. The R-symmetry under which these  central
charges transform is given  in Table \ref{centralcharges}. In
general dimensions the R-symmetry is given by SO$(d)$  but in $ D=4,3$
there is an extension of the R-symmetry to U(4)=SO(6)$\times$ SO(2)
and SO(8), respectively. This is in agreement with the fact that the duality symmetry
is enhanced to $\text{SL}(2,\mathbb{R}) \times \text{SO}(6,6)$ in $D=4$ and to $\text{SO}(8,8)$ in $D=3$.
In all cases the total number of charges,
including the translation generators, is $\tfrac{1}{2}( 16\times 17)
=136$.

The properties of the spinor charges in different dimensions
are as follows \cite{VanProeyen:1999ni}. In 10D the supercharges are
chiral Majorana spinors (16 components). In 9D and 8D they are
Majorana. In 7D one has a USp(2) doublet of spinors
satisfying symplectic Majorana conditions. In the non-chiral
six-dimensional theory (that we denote with 6A in Table
\ref{centralcharges}) one has R-symmetry $\text{USp}(2) \times
\text{USp}(2)$,  with the left-chiral supercharges in the $({\bf
2,1})$ and the right-chiral supercharges in the  $({\bf 1,2})$, and
each satisfying a symplectic Majorana condition. In the chiral (6B)
six-dimensional theory the chiral supercharge is in the ${\bf 4}$
of USp(4)  and satisfies a symplectic Majorana condition.
Similarly, in 5D one has a symplectic Majorana spinor in the ${\bf
4}$ of USp(4). In 4D the supercharge is a Majorana spinor in the
${\bf 4 } + \overline{\bf 4}$ of U(4). Finally, in 3D one has a
Majorana spinor in the ${\bf 8}_{\rm S}$ of SO(8).

\begin{table}[h!]
\begin{center}
\begin{tabular}{|c|c|c|c|c|c|c|c|}
\hline
$D$&$R$-symmetry&$n=0$&$n=1$&$n=2$&$n=3$&$n=4$&$n=5$\\[.1truecm]
\hline \rule[-1mm]{0mm}{6mm} 10&{\bf 1}&--&{\bf 1}&--&--&--&${\bf 1}^+$\\[.05truecm]
\hline \rule[-1mm]{0mm}{6mm} $\Delta$&&--&2&--&--&--&2\\[.05truecm]
\hline \rule[-1mm]{0mm}{6mm} 9&{\bf 1}&${\bf 1}$&{\bf 1}&--&--&${\bf 1}$&\\[.05truecm]
\hline \rule[-1mm]{0mm}{6mm} $\Delta$&&2&2&--&--&2,\,2&\\[.05truecm]
\hline \rule[-1mm]{0mm}{6mm} 8&U(1)& $ 2 \times {\bf 1}$ &{\bf 1} &
--& ${\bf 1}$
&${\bf 1}^+ + {\bf 1}^-$&\\[.05truecm]
\hline \rule[-1mm]{0mm}{6mm} $\Delta$&& 2 &2 &
--& 2,\,4 &$2+2$&\\[.05truecm]
\hline \rule[-1mm]{0mm}{6mm} 7&USp(2)& {\bf 3} & {\bf 1} & ${\bf 1}$ &{\bf 3} &&\\[.05truecm]
\hline \rule[-1mm]{0mm}{6mm} $\Delta$&& 2 & 2 & 2,\,9 &2,\,4&&\\[.05truecm]
\hline \rule[-1mm]{0mm}{6mm} 6A&USp(2)$\times$USp(2)&$({\bf 2},{\bf 2})$&$ ({\bf 1},{\bf 1})+  ({\bf 1},{\bf 1})$ & $({\bf 2},{\bf 2})$& $({\bf 3},{\bf 1})^+ + ({\bf 1},{\bf 3})^-$ &&\\[.05truecm]
\hline \rule[-1mm]{0mm}{6mm} $\Delta$&&2&$2+2,24$ & 2,\,10& $4+4$ &&\\[.05truecm]
\hline \rule[-1mm]{0mm}{6mm} 6B&USp(4)&--&${\bf 1} + {\bf 5}$ &--& ${\bf 10}^+$ &&\\[.05truecm]
\hline \rule[-1mm]{0mm}{6mm} $\Delta$&&--&$2+2,16$ &--&$ 4$&&\\[.05truecm]
\hline \rule[-1mm]{0mm}{6mm} 5&USp(4)& ${\bf 1} + {\bf 5}$& ${\bf 1} + {\bf 5} $& {\bf 10}&&&\\[.05truecm]
\hline \rule[-1mm]{0mm}{6mm} $\Delta$&& $2+2$&$2+ 2,32$& 4,\,12&&&\\[.05truecm]
\hline \rule[-1mm]{0mm}{6mm} 4&U(4)& ${\bf 6}+{\bf 6}$ & ${\bf 1} + {\bf 15}$ & ${\bf 10}^+ + {\bf \overline{10}}^-$ &&&\\[.05truecm]
\hline \rule[-1mm]{0mm}{6mm} $\Delta$&& $2+2$ &$3 + 4,64$ &$16+ 16$ &&&\\[.05truecm]
\hline \rule[-1mm]{0mm}{6mm} 3&SO(8)& ${\bf 28}_{\rm S}$ & ${\bf 1 } + {\bf 35}_{\rm S}$ &&&&\\[.05truecm]
\hline \rule[-1mm]{0mm}{6mm} $\Delta$&& 4 &$ 16 + 32 ,256$&&&&\\[.05truecm]
\hline
\end{tabular}
\end{center}
  \caption{\sl This table indicates the $R$-representations of the
  $n$-form central charges of $3\le D\le 10$ half-maximal supergravity and their relation to the half-supersymmetric heterotic branes.
   Momentum is included, corresponding to the always present $n=1$ singlet.
  If applicable, we have also indicated the space-time duality of
  the central charges with a superscript $\pm$. For each central charge the degeneracy $\Delta$ of the BPS conditions is indicated. The numbers after the comma refer to the degeneracy corresponding to the dual central charge.}
  \label{centralcharges}
\end{table}

 We now wish to determine which central charge corresponds to which
brane and how many branes correspond to a single central charge, i.e. how many branes have the same BPS condition. The number of branes that correspond to a single central charge is called the degeneracy $\Delta$ of the BPS condition in Table \ref{centralcharges}.
Remarkably, we find that for all heterotic branes,
provided we include  pp-waves and KK-monopoles, the relation between central charges and supersymmetric branes is given by the
following
\vskip .3truecm

\noindent {\bf central charge rule\,:}\ \
\begin{itemize}

\item
Given each  lightlike index $i\pm$, $ i=1,...,d$, of  $\text{SO}(d,d+n)$, interpret $i$
 as an  R-symmetry index. Then the resulting R-symmetry representation coincides with that of the relevant central charge. Due to the symmetry enhancement of the duality symmetry and R-symmetry
this identification requires that in $D=4$ the $\text{SL}(2,\mathbb{R})$ indices are converted to $\text{SO}(2)$ indices, while in $D=3$
the range of the indices is extended from 7 to 8.

\item The R-symmetry representation is simplified by applying the rule that, whenever a pair of two symmetric indices $ij$ of $\text{SO}(d)$
occur, this pair is replaced by the invariant tensor $\delta_{ij}$. This also applies to the $\text{SO}(2)$ part of the R-symmetry  in $D=4$, as well as for the $\text{SO}(8)$ R-symmetry in $D=3$.

\item If two branes of the same worldvolume dimension lead to the same R-symmetry representation for the charges using the rules above, then these branes correspond to the same central charge.
\end{itemize}

\bigskip

\noindent The above  rule also applies to branes and BPS conditions that
correspond to the dual central charges. In general all charges can be dualised, with the exception of the
0-form central charges and the translation generator.

The effect of the above central charge rule is that in all but three
exceptional cases, which will be discussed below\,\footnote{The
exceptions are 7D domain walls, 4D defect branes and 3D domain walls. The first case
is special since 7D domain walls are 5-branes
and there are 5-branes with hyper and tensor multiplets. The other two cases are
special because the 4D (3D) fundamental string is a defect brane (domain wall).},
the degeneracy of the BPS conditions for heterotic branes
is {\it twice} the degeneracy of the half-supersymmetric branes of
maximal supergravity\,\footnote{In this analysis we do not consider space-filling branes, since the degeneracy of these branes in the maximal case has not been discussed sofar in the literature. As we will see in the conclusions, the property that the degeneracy of the branes in the half-maximal theory is twice the degeneracy of those in the maximal theory continues to hold for space-filling branes, with one further exception of space-filling branes in six dimensions.}. For instance, according to the above rule the
standard heterotic branes, which couple to the fields $B_{1,A}$,
$B_2$, $D_{D-4}$ and $D_{D-3,A}$, have degeneracy 2 whereas the
standard branes of maximal supergravity have degeneracy 1. Starting
from 10D and going down in dimension  the first non-standard branes
are defect branes in 8D, which are 5-branes that couple to the
fields $D_{6,A_1A_2}$. According to the light-cone rule there are 4 of
them. They should be associated to the singlet $n=3$ charge and,
therefore, we find degeneracy 4. Similarly, we find that {\it  all}
heterotic defect branes have degeneracy 4 which is twice as much as
the degeneracy of the defect branes of maximal supergravity
\cite{Bergshoeff:2011se}. Domain walls first occur in 7D. In total
we have 9 domain walls, eight of them are solitonic and couple to
the fields $D_{6,A_1A_2A_3}$ and one of them is a $\alpha=-4$ domain
wall that couples to the field $F_6$. Although the domain walls occur in two different duality representations,
according to the central charge rule they correspond to the same R-symmetry representation, which is an SO(3) singlet,
and hence have the same BPS condition corresponding to the same singlet $n=5$ central charge.
We conclude that the  degeneracy $\Delta$ is 9.
This case is one of the three examples mentioned above where the degeneracy is not
just twice the degeneracy of the maximally supersymmetric case. This
has to do with the fact that this case involves two types of 5-branes,
one with hyper multiplets and one with tensor multiplets. Similarly, in maximal supergravity we have vector and tensor domain walls with different
degeneracies.

Below we verify the central charge rule for the different
dimensions, starting with 10D. We first note some general patterns. In any dimension the
translation generator corresponds to the pp-wave and the fundamental
string except in the 6B case where the fundamental string is
replaced by a KK monopole and in 3D where there is no pp-wave and, instead, we have the S-dual of the fundamental string.
 Furthermore, for $D\ge 5$ (in 6D we take
6A) there is always a singlet $n=D-5$ BPS condition corresponding to
the singlet solitonic brane and the KK monopole. Finally, as
mentioned above, the BPS conditions corresponding to the standard
branes (defect branes) always have a degeneracy $\Delta=2\ (\Delta=4)$.
\bigskip

\noindent {\sl 10D}\ \ In 10D the pp-wave and the fundamental string
have the same BPS condition corresponding to the $n=1$ translation generator.
Similarly, the heterotic 5-brane and the KK-monopole have
the same BPS condition corresponding to the $n=5$ central charge.
Note that the 5-form central charge is self-dual and
we do not consider the dual of the translation generator.
\bigskip

\noindent {\sl 9D}\ \ In 9D the fundamental 0-branes couple to the
fields $B_{1,\pm}$  and, therefore the $n=0$ BPS condition has
degeneracy 2. The fundamental string and pp-wave lead to a similar
degeneracy 2 of the $n=1$ BPS condition (the translation generator).
The 4-form central charge corresponds to the solitonic 4-brane and
the KK-monopole. Finally, the dual of the 4-form central charge
leads to a 5-form central charge which corresponds to the two
solitonic 5-branes that couple to $D_{6,\pm}$.
\bigskip

\noindent {\sl 8D}\ \ In 8D there are two $n=0$ BPS conditions. They
correspond to the 4 fundamental 0-branes that couple to $B_{1, i\pm}$
with $i=1,2$ a doublet of SO(2). The $n=1$ BPS condition corresponds
to the pp-wave and the fundamental string. The $n=3$ BPS condition
corresponds to a solitonic 3-brane and a KK-monopole. The two $n=4$
BPS conditions correspond  to the 4 solitonic 4-branes that couple
to $D_{5,i\pm}$. The supersymmetric
solitonic defect 5-branes that correspond to the single $n=5$ BPS
condition couple to the 4 solitonic fields $D_{6,i\pm j\pm}$ with $i \neq j$,
yielding a degeneracy 4 and a central charge that is a singlet of
SO(2) proportional to the invariant tensor $\epsilon_{ij}$.
\bigskip

\noindent {\sl 7D}\ \ The 3 $n=0$ BPS conditions correspond to the 6
fundamental 0-branes that couple to the fields $B_{1, i\pm}$ with
$i=1,2,3$ a vector of SO(3). Like before, the translation  generator
corresponds to the pp-wave and the fundamental string. The $n=2$ BPS
condition corresponds to the solitonic 2-brane and the KK-monopole.
The $n=3$ BPS conditions correspond to the 6 solitonic 3-branes that
couple to the fields $D_{4,i\pm}$.
The 3 $n=4$ BPS conditions correspond to defect branes and have
degeneracy 4. They correspond to the 12 defect branes that couple to
the fields $D_{5,1\pm2\pm}\,, D_{5,1\pm,3\pm}$ and $D_{5,2\pm3\pm}$.
Finally, the single $n=5$ BPS condition corresponds to the 8
solitonic domain walls that couple to the fields $D_{6,1\pm 2\pm
3\pm}$  and the single $\alpha=-4$ domain wall that couples to the
field $F_6$. Both fields correspond to the same central charge which
is a singlet of SO(3). The degeneracy is therefore 9. As explained
above, this case is exceptional due to the fact that we have two
types of domain walls, one with hyper multiplets and one with tensor multiplets.
\bigskip

\noindent {\sl 6A}\ \ The 4 $n=0$ BPS conditions correspond to the 8
fundamental 0-branes and have degeneracy 2, as expected. As usual,
the translation generator corresponds to the pp-wave and the
fundamental string. The second $n=1$ BPS condition corresponds to
the solitonic 1-brane and the KK-monopole.
The 4 $n=2$ BPS
conditions correspond to the 8 solitonic 2-branes and therefore have
degeneracy 2, as it should for standard branes. The corresponding
central charges are in the ${\bf 4}$ of SO(4). The 6 $n=3$ BPS
conditions correspond to defect branes with degeneracy 4. The
corresponding central charge transform as the self\-dual and
anti-selfdual representations of SO(4). The 4 $n=4$ BPS conditions
correspond to 32 solitonic domain walls that couple to $D_{5,1\pm
2\pm 3\pm}, D_{5, 1\pm 2\pm 4\pm}, D_{5,1\pm 3\pm 4\pm}$ and $D_{5,
2\pm 3\pm 4\pm}$ and 8 $\alpha=-4$ domain walls that couple to
$F_{5,i\pm}$. They both correspond to a central charge  in the ${\bf
4}$ of SO(4). This gives a total of 40 domain walls and hence
degeneracy $10$ which is twice as much as the degeneracy of domain
walls in non-chiral 6D maximal supergravity
\cite{Bergshoeff:2012pm}.  As Table \ref{centralcharges} shows, $D=6$ is the highest dimension in which there is an $n=1$ central charge other than momentum. This charge can be dualised, leading to an $n=5$ charge for space-filling branes. This charge couples to $D_{6, i_1\pm ...i_4\pm}$ and $F_{6, i\pm j\pm}$, which according to the central charge rule indeed all correspond to a singlet charge of $\text{SO}(4)$. The degeneracy is $16+8=24$.
\bigskip

\noindent {\sl 6B}\ \ This case is special since there is no dilaton
and hence no notion of $\alpha$ weight. The translation  generator
corresponds to a pp-wave and the KK monopole. The other 5 $n=1$ BPS
conditions correspond to the 10 strings that couple to $A_{2,i\pm}$.
This corresponds to a central charge in the ${\bf 5}$ of SO(5). The
10 $n=3$ BPS conditions correspond to the 40 defect  branes that
couple to $A_{4,i_1 \pm i_2 \pm}$ and hence have degeneracy $\Delta=4$. The central
charge is in the ${\bf 10}$ of SO(5). As in the 6A case, the $n=1$ charge in the ${\bf 5}$ can be dualised, giving an $n=5$ charge for $A_{6,  i\pm , j_1 \pm j_2 \pm}$. According to the central charge rule, the symmetric indices are proportional to $\delta_{i j}$ of SO(5), and one is left with a charge in the ${\bf 5} $ with degeneracy 16.

\bigskip

\noindent {\sl 5D}\ \ The 5+1 $n=0$ BPS conditions correspond to the
10 fundamental 0-branes that couple to $B_{1,i\pm}$, the solitonic
0-brane and the KK-monopole so that we have total degeneracy 2. The
translation generator corresponds to the pp-wave and the fundamental
string. The other 5 $n=1$ BPS conditions correspond to the 10
solitonic strings that couple to $D_{2,i\pm}$. These BPS conditions all
have degeneracy 2. The 10 $n=2$ BPS conditions correspond to the 40
defect branes that couple to $D_{3,i_1 \pm i_2 \pm}$ and have degeneracy 4.
The dual 10 $n=3$ BPS conditions correspond to 80 solitonic domain
walls that couple to $D_{4,i_1 \pm i_2 \pm i_3 \pm}$ and 40 $\alpha=-4$ domain
walls that couple to $F_{4,i_1\pm i_2 \pm}$. Both fields correspond to a
central charge in the ${\bf 10}$ of SO(5). This leads to a total of
120 domain walls and hence degeneracy $\Delta=12$ which is again twice as
much as the degeneracy of domain walls in 5D maximal supergravity
\cite{Bergshoeff:2012pm}. The $n=1$ charges in the ${\bf 5}$ can be dualised to $n=4$ charges, and indeed one can show that for all the space-filling branes in $D=5$ the central charge rule leads to a charge in the  ${\bf 5}$. The total degeneracy is 32.
\bigskip

\noindent {\sl 4D}\ \  The 12 $n=0$ BPS conditions correspond to the
12 fundamental 0-branes and 12 solitonic 0-branes. The fundamental
and solitonic branes have different BPS conditions since each of
them corresponds to a central charge in the ${\bf 6}$ of U(4) with a
{\sl different} U(1) weight. Note that this additional U(1) weight
occurs due to the fact that the R-symmetry is extended from SO(6) to
SO(6)$\times$SO(2). In 4D the singlet $n=1$ BPS condition
corresponds not only to the pp-wave and the fundamental string but
also to a $\alpha=-4$ 1-brane which  couples to $F_2$ and is the
S-dual of the fundamental string. We therefore have total degeneracy
3. Note that the fundamental string and its S-dual are defect branes.
This case is special due to the fact that these two defect branes have
the same BPS condition  as the pp-wave.  It is the second example
mentioned above
that violates the general central charge rule.
 The other 15 $n=1$ BPS conditions correspond to the 60 defect
1-branes that couple to $D_{2,i_1 \pm i_2 \pm}$. Unlike the defect branes
corresponding to the translation generator, these defect branes have
degeneracy $\Delta= 4$, characteristic for heterotic defect branes. The 20 $n=2$ BPS conditions correspond to 160 solitonic domain walls
that couple to $D_{3,i_1 \pm i_2 \pm i_3 \pm}$ and 160 $\alpha=-4$ domain walls
that couple to $F_{3, i_1 \pm i_2 \pm i_3 \pm}$ with central charges in the
self\-dual and anti-selfdual representation of SO(6). This leads to a
degeneracy $\Delta=16$ which is again twice as much as the degeneracy of
the supersymmetric domain walls of 4D maximal supergravity
\cite{Bergshoeff:2012pm}. Finally, one can dualise the $n=1$ charge in the ${\bf 15}$, giving an $n=3$ charge corresponding to the 960 space-filling branes associated to $D_{4, i_1\pm ...i_4 \pm}$ (240), $F_{4, i\pm, j_1\pm ...j_3\pm}$ (480) and $H_{4, i_1 \pm...i_4\pm}$ (240). The reader can check that all these branes are associated to the same BPS condition corresponding to a charge in the ${\bf 15}$ (that is the adjoint of $\text{SO}(6)$) by applying the central charge rule, thus resulting in a degeneracy $\Delta= 64$.

\bigskip

\noindent {\sl 3D}\ \ In 3D the 28 $n=0$ BPS conditions correspond
to defect branes. We have 14 fundamental defect 0-branes that couple
to $B_{1,i\pm}$ and 14 $\alpha=-4$ defect 0-branes that couple to
$F_{1,i\pm}$. They both correspond to a central charge in the ${\bf 7}$
of SO(7). Furthermore, we have 84 solitonic defect 0-branes that
couple to $D_{1,i_1 \pm i_2 \pm}$. They correspond to a central charge in the
${\bf 21}$ of SO(7). Together, this leads to a central charge in the
${\bf 28}_{\textrm{S}}$ of SO(8) with degeneracy 4, as we expect for heterotic
defect branes. The $n=1$ BPS conditions correspond to domain walls.
The singlet corresponds to one fundamental string, its S-dual (which couples to $J_2$ and replaces the pp-wave)
and  14 $\alpha=-4$ domain walls that couple to
$F_{2,i\pm j\pm}$. Note that the latter, according to the central charge rule, corresponds to a {\sl singlet} central charge.
Therefore, the singlet $n=1$ BPS condition  leads to
a degeneracy $\Delta=16$. This is the third special case mentioned above that is an exception to the general central charge rule.
The ${\bf 35}_{\rm S}$ $n=1$ BPS
conditions\,\footnote{Note that the ${\bf 35}_{\rm S}$ is a
self-dual representation of SO(8).} correspond to $8\times 35$
solitonic domain walls that couple to $D_{2,i_1 \pm i_2 \pm i_3 \pm}$, $16\times
35$ $\alpha=-4$ domain walls that couple to $F_{2, i_1 \pm ...i_4 \pm}$ and
$8\times 35$ domain walls that couple to $H_{2,i_1 \pm i_2 \pm i_3 \pm}$. This
leads to $32\times 35$ domain walls with degeneracy 32 which is
twice as much as in 3D maximal supergravity
\cite{Bergshoeff:2012pm}. We finally consider the space-filling branes, whose charge is the dual of the $n=1$ charge in the ${\bf 35}_{\rm S}$. The reader can check that all the space-filling branes, according to the central charge rule, give rise to charges as representations of $\text{SO}(7)$ that sum up to give the ${\bf 35}_{\rm S}$ of $\text{SO}(8)$. The degeneracy is $\Delta =  256$.  In all cases one can perform the same analysis considering directly representations of $\text{SO}(8,8+n)$ for the fields. In the particular case of the space-filling branes,
the branes correspond to $A_{3, \hat{i} \pm, \hat{j}_1 \pm ...\hat{j}_5 \pm}$ (where we denote with $\hat{i} \pm$  the light-cone indices of $\text{SO}(8,8+n)$) and applying the central charge rule one can see immediately that this gives a charge in the  ${\bf 35}_{\rm S}$.
\bigskip

This finishes our discussion about the relation between central charges and BPS
conditions plus their degeneracies.

\section{Conclusions}

In this work we have shown that the same brane wrapping rules we derived in our earlier work for the toroidally compactified Type IIA/IIB string theory, see Table \ref{wrappingrules}, also apply to the toroidally compactified heterotic theory. The heterotic wrapping rules are obtained from the ones given in Table \ref{wrappingrules} by restricting to the fundamental and solitonic branes only.
To derive the heterotic wrapping rules  we first classified the half-supersymmetric heterotic branes by requiring
a gauge-invariant and supersymmetric Wess-Zumino coupling or, equivalently, by picking out the real roots of the very extended $\text{SO}(8,8+n)$ algebra.
Here $n$ refers to the number of vector multiplets in ten dimensions. We next compared the numbers of such branes with the ten-dimensional ones and verified that they are connected by the wrapping rules given in Table \ref{wrappingrules}.
\bigskip

\begin{table}[h]
\begin{center}
\begin{tabular}{|c||c|c|c|c|c|}
\hline \rule[-1mm]{0mm}{6mm} type of brane &Fundamental& Dirichlet & Solitonic & E-branes & Space-filling\\
\hline \hline \rule[-1mm]{0mm}{6mm} wrapped&doubled&undoubled&undoubled&doubled&doubled\\
\hline \rule[-1mm]{0mm}{6mm} unwrapped&undoubled&undoubled&doubled&doubled&--\\
\hline
\end{tabular}
\caption{\sl \footnotesize The wrapping rules of the different types of IIA and IIB branes. The E-branes indicate branes with $\text{T}\sim g_s^{-3}$,
like the S-dual of the D7-brane. The last column indicates space-filling branes with $\text{T}\sim g_s^{-4}$ such as the S-dual of the D9-brane.
The heterotic wrapping rules are obtained by restricting  to the fundamental branes
($\text{T}\sim g_s^0$) and solitonic branes ($\text{T}\sim g_s^{-2}$) only.
\label{wrappingrules}}
\end{center}
\end{table}

We also discussed a so-called heterotic truncation of the IIA/IIB theory which projects the IIA/IIB branes onto the branes of the heterotic theory.
This rule can be understood as the restriction to those branes that have a common IIA and IIB origin. Sofar, we did not find an obvious generalization of this rule which applies to the full spectrum, including the mixed-symmetry fields,
of the  very extended Kac-Moody algebra $\text{E}_{11}$. We used the heterotic truncation to investigate the conjectured S-duality between
the heterotic theory on $T^4$ and the IIA theory on the orbifold realization $T^4/\mathbb{Z}_2$ of K3. We found that the S-duality between these two theories is consistent, at the level of the supersymmetric branes, with applying the same  wrapping rules we found for the toroidally compactified IIA and IIB theories, given in Table \ref{wrappingrules}, to the (even) cycles of the K3 orbifold.  We thus found for the first time that the wrapping rules also apply to (orbifold limits of) manifolds different from the torus. The fact that this result holds is not completely surprising  because
$T^4/\mathbb{Z}_2$ has 16 fixed points corresponding to 16 vector multiplets in the twisted sector, while the untwisted sector produces a symmetry $\text{SO}(4,4)$. This means that all the fields that in the heterotic theory are associated to branes according to the light-cone rule are dual to IIA fields coming from the untwisted sector of this orbifold.
It would be interesting to also study other orbifold limits of K3, i.e. $T^4 / \mathbb{Z}_n$ with $n=3,4,6$. It is not yet clear to us how to implement the wrapping rules in these cases.

Finally, we performed an in-depth
investigation of the relation between the central charges of the $D$-dimensional supersymmetry algebras with 16 supercharges and the
branes of the  $D$-dimensional heterotic theory. We established a simple so-called central charge rule which prescribes which T-duality
representation of heterotic branes is related to which R-symmetry representation of central charges. We found that in general the degeneracy of the heterotic BPS conditions, i.e.~how many independent branes satisfy the same BPS conditions, is twice as large as the degeneracies in the IIA/IIB theory. One can extend this analysis to include also the space-filling branes, whose degeneracy has not been discussed in the literature yet. By looking at the $n=1$ central charges of the maximal supersymmetry algebra which are different from the momentum operator (see Table 10 in \cite{Bergshoeff:2011zk}) and comparing this with the number of space-filling branes in various dimensions (see refs.~\cite{Bergshoeff:2011qk}, \cite{Kleinschmidt:2011vu} and \cite{Bergshoeff:2012ex}) one obtains the degeneracies which are summarised in Table \ref{maxsusydegeneracyspacefilling}. The reader may appreciate that in 5, 4 and 3 dimensions we again find that the  degeneracy of the space-filling branes in the half-maximal theories (see Table \ref{centralcharges}) is twice the degeneracy of the space-filling branes in the maximal theory. The six-dimensional case is an exception because in this case there are both tensor and vector branes. We have seen already in the previous section that the same exception to the rule occurs for the domain walls in seven dimensions. We hope to discuss in more detail the space-filling branes and their relation with the central charges for both the maximal and half-maximal theories in the near future.

\begin{table}[t]
\begin{center}
\begin{tabular}{|c|c|c|c|c|}
\hline
$D$&$R$-symmetry&$n=1$& space-filling branes& degeneracy\\[.1truecm]
\hline \rule[-1mm]{0mm}{6mm} IIA&{\bf 1}&{\bf 1}&0&0\\[.05truecm]
\hline \rule[-1mm]{0mm}{6mm} IIB&SO(2)&{\bf 2}&2&1\\[.05truecm]
\hline \rule[-1mm]{0mm}{6mm} 9&SO(2)&{\bf 2}&2&1\\[.05truecm]
\hline \rule[-1mm]{0mm}{6mm} 8&U(2)&{\bf 3} & 6 & 2\\[.05truecm]
\hline \rule[-1mm]{0mm}{6mm} 7&Sp(4) & {\bf 5} & 20 & 4\\[.05truecm]
\hline \rule[-1mm]{0mm}{6mm} 6&Sp(4)$\times$Sp(4)&$({\bf 1},{\bf 1})$ & 16& 16\\[.05truecm]
\rule[-1mm]{0mm}{6mm} & &$({\bf 1},
{\bf 5}) +({\bf 5},
{\bf 1}) $ &80 &8\\[.05truecm]
\hline \rule[-1mm]{0mm}{6mm} 5&Sp(8)& ${\bf 27}$ & 432&16\\[.05truecm]
\hline \rule[-1mm]{0mm}{6mm} 4&SU(8)& {\bf 63} & 2016 & 32\\[.05truecm]
\hline \rule[-1mm]{0mm}{6mm} 3&SO(16)& {\bf 135} &17280 & 128\\[.05truecm]
\hline
\end{tabular}
\end{center}
  \caption{\sl  In this table we determine the degeneracy of the space-filling branes of the maximal theories with respect to the $n=1$ central charges in all dimensions. In the six-dimensional case, the first line corresponds to tensor branes and the second line to vector branes.}
  \label{maxsusydegeneracyspacefilling}
\end{table}

The fact that the wrapping rules given in Table \ref{wrappingrules} apply both to the toroidally compactified IIA, IIB and  heterotic theories
and, furthermore, also apply to the K3 orbifold, is encouraging. It suggests that the wrapping rules give a hint about the geometry underlying the full non-perturbative string theory. Restricting to the perturbative fundamental branes our wrapping rules are in line with the doubled geometry proposal \cite{Hull}\,\footnote{Note that we do not consider here an extension of space-time itself, like in e.g.~\cite{West:2003fc} or \cite{Hull:2009mi}.
}. Indeed, the doubling upon wrapping means that the fundamental string effectively sees a doubled torus.
An alternative interpretation is that there
is a single torus and that the doubling upon wrapping is due to the presence of an extra object in ten dimensions, i.e.~the pp-wave, which upon torus reduction leads to the desired doubling of wrapped strings. It should be stressed  that the worldvolume action of the fundamental branes,
always contain twice as many transverse embedding scalars as compactified directions as required by the $\text{SO}(d,d)$ T-duality. On the other
hand, the background fields that occur in the Wess-Zumino coupling to the branes that we have been studying depend only the usual spacetime coordinates.

The situation becomes more subtle if we include, in the IIA/IIB case, the D-branes as well. According to Table \ref{wrappingrules} no doubling upon wrapping takes place
or, in other words, there is no Dirichlet analogue of the pp-wave. This means, for instance, that the  D-string, unlike the fundamental string, does not see a doubled torus even though its worldvolume action does contain the same doubled number of embedding scalars as the fundamental string.
Proceeding to the solitonic branes, we see from Table \ref{wrappingrules} that these branes are governed by a so-called dual wrapping rule.
This dual rule prescribes that the number of solitonic branes is doubled when un-wrapped instead of wrapped. It is hard to understand the
doubling upon un-wrapping from a doubled geometry perspective only. Alternatively, the doubling upon un-wrapping, can be understood from the fact that string theory contains Kaluza-Klein monopoles that upon toroidal reduction leads to the desired doubling. The tricky thing with this interpretation is that we found that the dual wrapping rule also applies to solitonic branes with 2 or less transverse directions. To realize the dual wrapping rule
for these kind of branes requires a generalization of Kaluza-Klein monopoles to generalised monopoles with 2 or less transverse directions. At the moment it is
not clear whether such objects can be defined within string theory. Scanning the remaining branes in ten dimensions, see Table \ref{wrappingrules}, we obtain further wrapping rules whose interpretation in terms of a doubled geometry is not clear.

What we find is that all  branes with a fixed dilaton scaling of the tension, i.e.~those branes that are related to each other by a perturbative symmetry,
see the same kind of geometry. However, branes with a different dilaton scaling of the tension see a different kind of geometry. The doubled geometry
occurs in the case of the fundamental branes. In this context we remind that the Type I string, which may be obtained from a so-called Type I truncation of the IIB theory (which is the low-energy manifestation of the orientifold projection \cite{Sagnotti:1987tw})  and which is non-perturbative from the heterotic point of view, sees a quite different geometry than the heterotic string. The heterotic wrapping rules do not apply to the Type I branes.
This is to be expected  because
the type I theory describes unoriented closed strings and open strings, while  the heterotic wrapping rules are a manifestation of the fact that the strings are closed and oriented.

A further understanding of how to interpret the different wrapping rules we found is needed. They give a clue about what the geometry is that
is seen by the different branes of string theory.
We hope to come back to this issue in the nearby future.

\section*{Acknowledgements}

This work was initiated at the Isaac Newton Institute for
Mathematical Sciences (Cambridge). We thank the organizers of the ``Mathematics
and Applications of Branes in String and M-theory'' programme for
their hospitality and financial support. FR would like to thank G.~Pradisi and L.~Romano
for useful discussions.

\vskip 1.5cm

\appendix

\section{\label{appendix1}The ${SO}(8, 8+n)^{+++}$ Kac-Moody spectrum}

In this appendix we want to obtain the half-supersymmetric branes of the heterotic theory  from an analysis of the roots of the Kac-Moody algebra $\text{SO}(8, 8+n)^{+++}$, that is the very-extended $\text{SO}(8,8+n)$ algebra. This case is technically different from the maximal case corresponding to the Kac-Moody algebra $\text{E}_{8(8)}^{+++}$, which is the very-extended  $\text{E}_{8(8)}$ algebra. The reason for this is that whereas $\text{E}_{8(8)}$ is maximally non-compact,  i.e.~in split form,  the algebra $\text{SO}(8,8+n)$ is  in split form  only for $n=0$, 1 and $-1$.~\footnote{The $n=-1$ case corresponds to pure half-maximal supergravity in nine dimensions, and cannot be uplifted to ten dimensions. We will not consider this case here because we are only interested in theories that can be uplifted to ten dimensions.}.  We know  that in  the split case  all real roots, i.e.~the ones with  squared length $\alpha^2=2$, correspond to fields associated to half-supersymmetric branes in the theory \cite{Kleinschmidt:2011vu}. Extending this rule to the non-split case requires
a proper definition of real roots for the non-split case.

The analysis of the forms resulting from  $\text{SO}(8, 8+n)^{+++}$ in any dimension was performed in  \cite{Bergshoeff:2007vb}  (see Tables 2 and  3 of that paper for a summary of results). Here we will refine this analysis by specifying  the squared length of the corresponding root. In order to study the reality properties of the roots, one has to specify the real form of the algebra. This can be done  by means of the so-called Tits-Satake diagrams. For a detailed analysis of a Tits-Satake diagram, see for instance the review \cite{Henneaux:2007ej} and references therein. Here it is enough to mention that a Tits-Satake diagram is a Dynkin diagram where the nodes contain the following additional information:
\begin{itemize}
\item to each imaginary root, that is a root fixed under the Cartan involution,~\footnote{The Cartan involution is an involution that makes the Killing form negative-definite. Therefore, for a compact real form  the Cartan involution is simply the identity.}  one associates a painted node;

\item to each real simple root one associates an unpainted node;

\item to each two complex simple root orbit under the Cartan involution one draws an arrow joining them.
\end{itemize}

\begin{figure}[h]
\begin{center}
\begin{picture}(340,20)

\put(10,10){\circle{10}}
\put(50,10){\circle{10}}
\put(15,10){\line(1,0){30}}
\put(55,10){\line(1,0){10}}
\put(80,10){\line(1,0){10}}
\put(100,10){\line(1,0){30}}
\put(95,10){\circle{10}}
\put(135,10){\circle*{10}}
\put(140,10){\line(1,0){5}}
\put(155,10){\line(1,0){15}}
\put(180,10){\line(1,0){30}}
\put(175,10){\circle*{10}}
\put(215,10){\circle*{10}}
\put(255,10){\circle*{10}}
\put(218,13){\line(1,0){34}}
\put(218,7){\line(1,0){34}}
\put(240,10){\line(-1,1){10}}
\put(240,10){\line(-1,-1){10}}

\put(300,10){$\text{SO}(2n-p+1,p)$}
\put(307,-10){$0\leq p \leq n$}

\put(8,-8){$1$}
\put(48,-8){$2$}
\put(90,-8){$p$}
\put(120,-8){$p+1$}
\put(158,-8){$n-2$}
\put(200,-8){$n-1$}
\put(253,-8){$n$}

\end{picture}
\vskip .4truecm
\caption{\sl \footnotesize The Tits-Satake diagrams corresponding to the real forms of $\text{B}_n$. All nodes from $p+1$ to $n$ are painted. The case $p$, in which  all nodes are unpainted, corresponds to the split form $\text{SO}(n+1,n)$. \label{TitsSatakeBn}}
\end{center}
\end{figure}
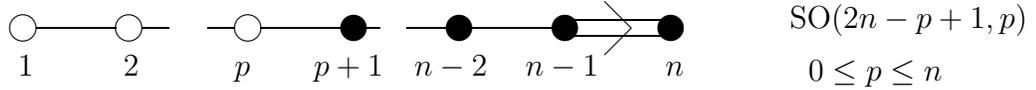

We have drawn the Tits-Satake diagrams for the various real forms of the algebras $\text{D}_n$ and $\text{B}_n$ in Figs.~\ref{TitsSatakeDn} and  \ref{TitsSatakeBn}.

\begin{figure}[h]
\begin{center}
\begin{picture}(340,200)
\put(10,170){\circle{10}}
\put(50,170){\circle{10}}
\put(90,170){\circle{10}}
\multiput(15,170)(40,0){2}{\line(1,0){30}}
\put(95,170){\line(1,0){15}}
\put(125,170){\line(1,0){15}}
\put(155,170){\line(1,0){15}}
\put(180,170){\line(1,0){30}}
\put(175,170){\circle{10}}
\put(215,170){\circle{10}}
\put(218,173){\line(1,1){20}}
\put(218,167){\line(1,-1){20}}
\put(241,196){\circle{10}}
\put(241,144){\circle{10}}
\put(300,170){$\text{SO}(n,n)$}
 \put(8,152){$1$}
\put(48,152){$2$}
\put(88,152){$3$}
\put(158,152){$n-3$}
\put(197,152){$n-2$}
\put(250,140){$n-1$}
\put(250,193){$n$}

\put(10,90){\circle{10}}
\put(50,90){\circle{10}}
\put(90,90){\circle{10}}
\multiput(15,90)(40,0){2}{\line(1,0){30}}
\put(95,90){\line(1,0){15}}
\put(125,90){\line(1,0){15}}
\put(155,90){\line(1,0){15}}
\put(180,90){\line(1,0){30}}
\put(175,90){\circle{10}}
\put(215,90){\circle{10}}
\put(218,93){\line(1,1){20}}
\put(218,87){\line(1,-1){20}}
\put(241,116){\circle{10}}
\put(241,64){\circle{10}}
\put(241,75){\line(0,1){30}}
\put(241,75){\line(1,1){10}}
\put(241,75){\line(-1,1){10}}
\put(241,105){\line(-1,-1){10}}
\put(241,105){\line(1,-1){10}}
\put(300,90){$\text{SO}(n+1,n-1)$}
 \put(8,72){$1$}
\put(48,72){$2$}
\put(88,72){$3$}
\put(158,72){$n-3$}
\put(197,72){$n-2$}
\put(250,60){$n-1$}
\put(250,113){$n$}

\put(10,10){\circle{10}}
\put(50,10){\circle{10}}
\put(15,10){\line(1,0){30}}
\put(55,10){\line(1,0){10}}
\put(80,10){\line(1,0){10}}
\put(100,10){\line(1,0){30}}
\put(95,10){\circle{10}}
\put(135,10){\circle*{10}}
\put(140,10){\line(1,0){5}}
\put(155,10){\line(1,0){15}}
\put(180,10){\line(1,0){30}}
\put(175,10){\circle*{10}}
\put(215,10){\circle*{10}}
\put(218,13){\line(1,1){20}}
\put(218,7){\line(1,-1){20}}
\put(241,36){\circle*{10}}
\put(241,-16){\circle*{10}}
\put(300,10){$\text{SO}(2n-p,p)$}
\put(305,-10){$0\leq p \leq n-2$}

\put(8,-8){$1$}
\put(48,-8){$2$}
\put(90,-8){$p$}
\put(120,-8){$p+1$}
\put(158,-8){$n-3$}
\put(197,-8){$n-2$}
\put(250,-20){$n-1$}
\put(250,33){$n$}

\end{picture}
\vskip .4truecm
\caption{\sl \footnotesize The Tits-Satake diagrams corresponding to the real forms $\text{SO}(2n-p,p)$ of $\text{D}_n$. We are not considering here the $\text{SO}^*(2n)$ real form because it is not relevant for our analysis. In the last diagram, all nodes from $p+1$ to $n$ are painted. \label{TitsSatakeDn}}
\end{center}
\end{figure}
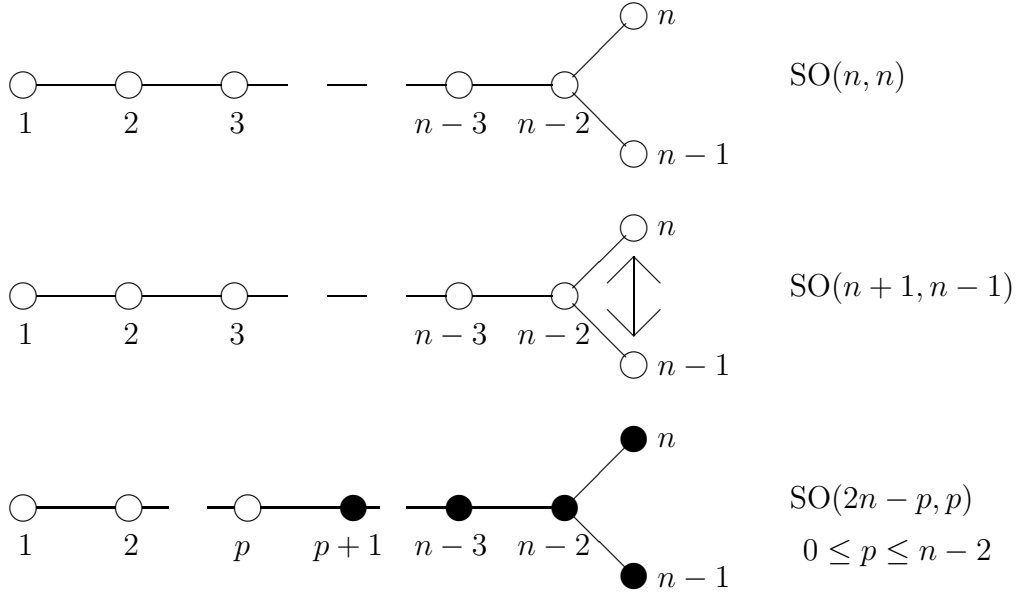

We  consider first the maximally non-compact cases $\text{SO}(n,n)$ and $\text{SO}(n+1,n)$, in which case all the simple roots can be taken to be real.  In the $\text{SO}(n,n)$ case, denoting with $i\pm$, $i=1,...,n$ the lightlike directions (as we have done throughout the paper), one can consider the generators corresponding to the simple roots as follows:
  \begin{equation}
  \alpha_1 \rightarrow T_{1+ \ 2-} \qquad \alpha_2 \rightarrow T_{2+\  3-} \qquad ...\quad \alpha_{n-1} \rightarrow T_{n-1+\  n-} \qquad \alpha_n \rightarrow T_{n-1 +\  n+ } \quad  .\label{rootgeneratorSOnn}
\end{equation}
From this set of generators, using the fact that the indices are contracted using the symmetric invariant tensor
  \begin{equation}
  {\mathbb I}_{i+\  j-} =   {\mathbb I}_{i-\  j+} = \delta_{ij} \quad , \quad   {\mathbb I}_{i+\  j+} =   {\mathbb I}_{i- \ j-}  =0 \quad ,
\end{equation}
one recovers the whole set of positive roots by constructing all possible contractions of the tensors above~\footnote{The same applies to the negative roots, using the rule that the generators associated to the negative roots are given by changing both light-cone signs of the indices of the generator associted to the corresponding positive roots.}.  All the roots have the same length, $\alpha^2 =2$. In the $\text{SO}(n+1,n)$ case, denoting with 1 the single spacetime index and again with  $i\pm$, $i=1,...,n$ the lightlike directions, the correspondence between roots and generators is
\begin{equation}
  \alpha_1 \rightarrow T_{1+ \ 2-} \qquad \alpha_2 \rightarrow T_{2+ \ 3-} \qquad ...\quad \alpha_{n-1} \rightarrow T_{n-1+\ n-} \qquad \alpha_n \rightarrow T_{ n + \ 1 } \quad , \label{rootgeneratorSOnn+1}
\end{equation}
where the last root $\alpha_n$  is the short ($\alpha^2 =1$) simple root. In this case the symmetric  invariant tensor is as before with the addition of ${\mathbb I}_{ 1 \ 1 } = 1$ that contracts the index in the spacelike direction, and one obtains all the positive roots as sums of simple roots by contracting in all possible ways the generators above.
For any $n$, this algebra contains $n$ short ($\alpha^2 =1$)  positive roots, which are associated to the generators $T_{i+ \ 1}$. This can be seen by acting recursively on $T_{n+ \ 1}$ with the other generators in eq.~\eqref{rootgeneratorSOnn+1}.  All the other roots have $\alpha^2 =2$.

When one considers different real forms, one can define the generators exactly in the same way, but clearly now the definition of light-cone directions is on the complex numbers. The reader can check that, if one defines the generators  as in eqs.~\eqref{rootgeneratorSOnn} and \eqref{rootgeneratorSOnn+1},  for any $p$ the generators that become imaginary are precisely in correspondence with the imaginary roots of $\text{SO}(2n-p,p)$ and $\text{SO}(2n-p+1,p)$ as dictated by the Tits-Satake diagrams given in Figs. \ref{TitsSatakeDn} and \ref{TitsSatakeBn}. This means that in general the generators that correspond to the real $\alpha^2=2$ roots are the ones along the lightlike directions and satisfy the light-cone rule used in this paper. For instance, as a trivial example one can consider the compact cases $\text{SO}(2n)$ and $\text{SO}(2n+1)$, in which case all roots are imaginary and correspondingly there are no lightlike directions.
Moreover, the same applies to the weights: the Tits-Satake diagrams naturally give a dictionary for the reality properties of the weights, and translating this to the corresponding representations $O_{A_1 ...A_p , B_1 ...B_q ,...}$ one can verify, using the generators above, that the real weights are associated to the directions satisfying the light-cone rule.~\footnote{This can be generalised easily to representations  containing spinorial indices. The light-cone rule extends to these representations \cite{Bergshoeff:2011ee}. We did not consider this extension in this paper since spinorial indices do not occur in the heterotic case.}

In  \cite{Riccioni:2008jz} it was shown how one can define very-extended versions of real algebras that are not in the split form using the Tits-Satake diagrams. From that analysis,
it naturally follows that the theory corresponding to the Kac-Moody algebra $\text{SO}(8,8+n)^{+++}$, for any $n \geq 0$,  can only be uplifted up to 10 dimensions.  It also follows naturally from the very-extended version of the Tits-Satake diagram that the internal symmetry of the $D$-dimensional theory is $\text{SO}(d,d+n)$ for $D\geq 5$, $\text{SL}(2,\mathbb{R}) \times \text{SO}(6,6+n)$ in 4D and $\text{SO}(8,8+n)$ in 3D.  In order to determine the components of the T-duality representations of the fields that correspond to branes in any dimension, one proceeds as follows. One decomposes the adjoint of $\text{SO}(8,8+n)^{+++}$ in representations of $\text{GL}(D,\mathbb{R})\times \text{SO}(d,d+n)$, and only considers the representations of $\text{GL}(D,\mathbb{R})$ having $p$ antisymmetric indices (corresponding to $p$-forms). One then selects only the representations whose highest weight is associated to a real root of $\text{SO}(8,8+n)^{+++}$ with squared-length $\alpha^2=2$ (this can for instance be done using the programme SimpLie~\cite{Bergshoeff:2007qi}). Within such representations, one then uses the analysis above, which selects all components that are associated to real roots of $\text{SO}(8,8+n)^{+++}$ as the ones that satisfy the light-cone rules of  $ \text{SO}(d,d+n)$. This shows that the WZ analysis in section 2 and the analysis of $\alpha^2=2$ roots give the same answer also in the half-maximal case.

As a corollary, we observe that the squared-length of the roots of  $\text{SO}(n,m)^{+++}$ satisfy a universal pattern which is exactly in agreement with the analysis above. The pattern is the following. In $D$ dimensions, one  decomposes  $\text{SO}(n,m)^{+++}$ in $\text{GL}(D,\mathbb{R})\times \text{SO}(n-D+2, m-D+2)$.~\footnote{In $D=3$ one has symmetry enhancement to $\text{SO}(n,m)$, and in $D=4$ to $\text{SL}(2,\mathbb{R})\times \text{SO}(n-2,m-2)$. Similarly, there is an additional possible six-dimensional decomposition for $D=6$, giving $\text{SO}(n-3,m-3)$. We have seen all this in detail throughout the paper.} The forms, that are antisymmetric representations of $\text{GL}(D,\mathbb{R})$, have a universal structure as representations of $\text{SO}(n-D+2, m-D+2)$, which does not depend on $n$ and $m$. It is convenient to introduce the notation $q =
n+m-2D+4$.
Now take $q$ large enough and start reducing it unit by unit and determine in each case the squared-length of the roots associated to the highest weights of the representations. Consider in particular a $p$-form  representation of  $\text{SO}(n-D+2, m-D+2)$ with $r$ antisymmetric indices $A_{p,A_1 ...A_r}$. One may always use the epsilon symbol $\epsilon_{A_1 ...A_q}$   to convert $r$ indices into $q-r $ indices. As soon as $q < 2r$, you decrease the number of indices by doing this. Correspondingly,
 when this happens, the squared length $\alpha^2$ decreases by the amount
 \begin{equation}
\Delta \alpha^2 = q -2r \quad . \label{newalphasquared}
\end{equation}
This is exactly in agreement with our light-cone analysis above. Consider as an example the split case $m$. In this case $q$ is even and the algebra is given by $\text{SO}(q/2,q/2)$.  Suppose that the $p$-form $A_{p,A_1 ...A_r}$ has $\alpha^2 =2$ for $q \geq 2r$. If you decrease $q$, as soon as $q< 2r$ there are no longer components of the representation that satisfy the light-cone rule. From eq.~\eqref{newalphasquared}, we see that the value of $\alpha^2$ decreases accordingly and the highest weight no longer corresponds to a real root.

\section{Type I Truncation}

\begin{figure}[h]
\begin{center}
\begin{picture}(340,70)
\multiput(10,10)(40,0){9}{\circle{10}}
\multiput(15,10)(40,0){8}{\line(1,0){30}}
\multiput(135,10)(40,0){1}{\line(1,0){30}} \put(130,50){\circle{10}}
\put(130,15){\line(0,1){30}}
 \put(290,50){\circle{10}}
\put(290,15){\line(0,1){30}}
 \put(8,-8){$1$}
\put(48,-8){$2$}
\put(88,-8){$3$}
\put(128,-8){$4$}
\put(168,-8){$5$}
\put(208,-8){$6$}
\put(248,-8){$7$}
\put(288,-8){$8$}
\put(328,-8){$9$}
 \put(140,47){$10$}
 \put(300,47){$11$}
\end{picture}
\vskip .4truecm
\caption{\sl \footnotesize The $\text{SO}(8,8)^{+++}$ Dynkin diagram. \label{D8+++}}
\end{center}
\end{figure}
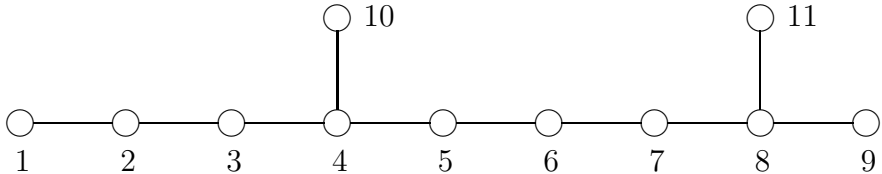

 In this appendix we consider the truncation to the low-energy effective action of the closed sector of the Type I string theory. Unlike the heterotic case, the Type I theory can only be obtained by a truncation of the IIB theory. This can be easily understood by comparing the supergravity sector of the Heterotic and Type I spectrum. Both result from  the $\text{SO}(8,8)^{+++}$ diagram, see Figure \ref{D8+++}. From the diagram, deleting nodes 10 and 11, one obtains the 10 dimensional spectrum of the ${\cal N}=1$ theory with no vector multiplets. Denoting with $(l_{10},l_{11})$ the levels corresponding to the two deleted nodes, we get the spectrum of forms (the last number in brackets denotes the squared length of the corresponding root)
 \begin{eqnarray}
   & & (0,1): \ \ \qquad \qquad  A_2 \ \ (2)\,, \nonumber \\
   & & (1,0): \ \ \qquad \qquad  A_6 \ \ (2)\,, \nonumber \\
   & & (1,1): \ \  \qquad \qquad A_8 \ \ (0)\,, \nonumber \\
   & & (1,2): \ \  \qquad \qquad A_{10} \ \ (-2) \quad .
\end{eqnarray}

To obtain the heterotic theory, one assigns a dilaton scaling
  \begin{equation}
\alpha_{\rm Het} = - 2 l_{10} \quad . \label{alphaheterotic}
\end{equation}
Given that the node $l_{10}$ only  enters the internal symmetry after compactification to 4D and 3D,  the heterotic internal symmetry is perturbative
for $D>4$, i.e.~it does not involve the dilaton. In 4D there is an extra $\text{SL}(2,\mathbb{R})$ (indeed node 10) that involves the dilaton, and  in 3D the $\text{SO}(7,7)$ T-duality symmetry, that does not transform the dilaton, is enhanced to the non-perturbative symmetry $\text{SO}(8,8)$.

To obtain the Type I theory, one assigns a different dilaton scaling
  \begin{equation}
\alpha_{\rm type-I} = - l_{10} - l_{11} \label{alphatypeI}
 \quad .
\end{equation}
The difference is that while in the heterotic case $A_2$ is fundamental and $A_6$ is solitonic, in the Type I case both $A_2$ and $A_6$
are Dirichlet. Whereas both the IIA and the IIB theory contain a fundamental 2-form and thus can be both truncated to the heterotic theory, only the IIB theory contains a Dirichlet 2-form as well.
Therefore, the closed sector of the Type I theory can only be obtained by a truncation of the IIB theory.  From eqs.~\eqref{alphaheterotic} and \eqref{alphatypeI} it also follows that
 \begin{equation}
\alpha_{\rm type-I} = -\frac{1}{2} p - \alpha_{\rm Het} \quad ,\label{hetversustypeI}
\end{equation}
relating the two dilaton scalings $\alpha_{\rm type-I}$ and $\alpha_{\rm Het}$ for each $p$-form.

\begin{figure}[h]
\begin{center}
\begin{picture}(380,70)
\multiput(10,10)(40,0){10}{\circle{10}}
\multiput(15,10)(40,0){9}{\line(1,0){30}}
\multiput(135,10)(40,0){1}{\line(1,0){30}}
 \put(290,50){\circle{10}}
\put(290,15){\line(0,1){30}}
 \put(8,-8){$1$}
\put(48,-8){$2$}
\put(88,-8){$3$}
\put(128,-8){$4$}
\put(168,-8){$5$}
\put(208,-8){$6$}
\put(248,-8){$7$}
\put(288,-8){$8$}
\put(328,-8){$9$}
 \put(368,-8){$10$}
 \put(300,47){$11$}
\end{picture}
\vskip .4truecm
\caption{\sl \footnotesize The $\text{E}_8^{+++}$ Dynkin diagram. \label{E8+++}}
\end{center}
\end{figure}
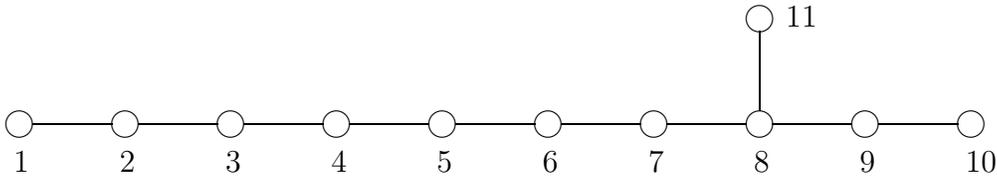

One can generalise the IIB truncation to  Type I at the level of the full Kac-Moody algebra
$\text{E}_8^{+++}$ , whose Dynkin diagram is given in Fig. \ref{E8+++}, exactly as we did in Subsection 3.2 for the Heterotic theory. In the diagram of Fig. \ref{E8+++},
the IIA theory corresponds to deleting nodes 10 and 11, while the IIB theory corresponds to deleting 9 and 10.  Denoting with $m_{10}$ the  level of node 10 of the $\text{E}_{8}^{+++}$ diagram, one has for both IIA and IIB
 \begin{equation}
\alpha_{\rm IIA/IIB} = -m_{10} \quad .
\end{equation}
We have seen in Subsection 3.2 that
 both theories can be truncated to the heterotic theory, and  the prescription  in both cases is to truncate to even $m_{10}$, that is even $\alpha$, and then project away additional fields (if one is interested in form fields after dimensional reduction to six dimensions and above, these extra fields are given in  eq.~\eqref{projectedoutallfieldsIIA} for IIA and in   eq.~\eqref{projectedoutallfieldsIIB} for IIB).  The dilaton scaling  leads to the identification
\begin{equation}
 2 l_{10} = m_{10} \quad .
\end{equation}

The Type I truncation is obtained by taking the fields in the IIB theory with
$m_9 + m_{10}$ even (like in the heterotic truncation, this does not mean that we keep {\it all} such fields). In this case, to match the dilaton scaling, the identification is
  \begin{equation}
l_{10} + l_{11} = m_{10} \quad .
\end{equation}

The fact that the dilaton scaling in the Type-I truncation involves  node $l_{11}$  implies that in the Type I case the internal symmetry is non-perturbative in any dimension. For instance, in nine dimensions the $\text{SO}(1,1)$ vector of 1-forms comes from the graviton, that is $\alpha_{\rm type-I} = 0$, and from the reduced R-R 2-form, with $\alpha_{\rm type-I} = -1$\,\footnote{Note that in the heterotic case the 2-form, like the graviton, has $\alpha_{\rm het}=0$.}. In general,
fields with different $\alpha$'s are involved in building up  representations of $\text{SO}(d,d)$. This implies that, unlike in the heterotic case, the truncation does not preserve the wrapping rule. This can already be seen from the nine-dimensional example above.

The Kac-Moody analysis of the Type I spectrum can be extended to the full algebra $\text{SO}(8,8+n)^{+++}$ to include the matter sector. From
eq.~\eqref{hetversustypeI} one can see that in ten dimensions the 1-forms $B_{1,A}$ in the fundamental of $\text{SO}(n)$ in the heterotic theory are mapped to 1-forms with $\alpha_{\rm type-I}= -\frac{1}{2}$, while the dual forms $D_{7,A}$ are mapped to 7-forms with $\alpha_{\rm type-I}= -\frac{3}{2}$ . These half-integer  dilaton scalings are related to the fact that this part of the spectrum of the theory comes from the open sector.


\begin{thebibliography}{99}



\bibitem{book} For an introduction into supergravity, see, e.g.,
T.~Ort\'in, ``Gravity and Strings'', Cambridge University Press (2007), ISBN 0521035465 and ``Supergravity'', Daniel Z. Freedman and Antoine Van Proeyen, Cambridge University Press (2012), ISBN 0521194016.

\bibitem{Lu:1997bg}
For an early reference, see, e.g.,   H.~Lu, C.~N.~Pope and K.~S.~Stelle,
  ``Multiplet structures of BPS solitons,''  Class.\ Quant.\ Grav.\  {\bf 15} (1998) 537  [hep-th/9708109].  


\bibitem{Bekaert:2002uh}
See. e.g.,  X.~Bekaert, N.~Boulanger and M.~Henneaux,
  ``Consistent deformations of dual formulations of linearized gravity: A No go result,''  Phys.\ Rev.\ D {\bf 67} (2003) 044010  [hep-th/0210278];  
  E.~A.~Bergshoeff, M.~de Roo, S.~F.~Kerstan, A.~Kleinschmidt and F.~Riccioni,
  ``Dual Gravity and Matter,''  Gen.\ Rel.\ Grav.\  {\bf 41} (2009) 39  [arXiv:0803.1963 [hep-th]].  


\bibitem{West:2001as}
  P.~C.~West,
  ``E(11) and M theory,''  Class.\ Quant.\ Grav.\  {\bf 18} (2001) 4443  [hep-th/0104081].  


\bibitem{West:2003fc}
  P.~C.~West,
  ``E(11), SL(32) and central charges,''  Phys.\ Lett.\ B {\bf 575} (2003) 333  [hep-th/0307098].  


\bibitem{Bergshoeff:2011ee}
  E.~A.~Bergshoeff and F.~Riccioni,
  ``Branes and wrapping rules,''  Phys.\ Lett.\ B {\bf 704} (2011) 367  [arXiv:1108.5067 [hep-th]];  
  E.~Bergshoeff and F.~Riccioni,
  ``Branes and wrapping rules,''  J.\ Phys.\ Conf.\ Ser.\  {\bf 343} (2012) 012015.  


\bibitem{de Azcarraga:1989gm}
  J.~A.~de Azcarraga, J.~P.~Gauntlett, J.~M.~Izquierdo and P.~K.~Townsend,
  ``Topological Extensions of the Supersymmetry Algebra for Extended Objects,''  Phys.\ Rev.\ Lett.\  {\bf 63} (1989) 2443.  


\bibitem{Townsend:1997wg}
  P.~K.~Townsend,
  ``M theory from its superalgebra,''  In *Cargese 1997, Strings, branes and dualities* 141-177  [hep-th/9712004].  

\bibitem{Bergshoeff:2012pm}
  E.~A.~Bergshoeff, A.~Kleinschmidt and F.~Riccioni,
  ``Supersymmetric Domain Walls,''  arXiv:1206.5697 [hep-th].  

\bibitem{Schnakenburg:2004vd}
  I.~Schnakenburg and P.~C.~West,
  ``Kac-Moody symmetries of ten-dimensional nonmaximal supergravity theories,''  JHEP {\bf 0405} (2004) 019  [hep-th/0401196].  



\bibitem{Bergshoeff:2007vb}
  E.~A.~Bergshoeff, J.~Gomis, T.~A.~Nutma and D.~Roest,
  ``Kac-Moody Spectrum of (Half-)Maximal Supergravities,''  JHEP {\bf 0802} (2008) 069  [arXiv:0711.2035 [hep-th]].  

\bibitem{Bergshoeff:2010xc}
  E.~A.~Bergshoeff and F.~Riccioni,
  ``D-Brane Wess-Zumino Terms and U-Duality,''  JHEP {\bf 1011} (2010) 139  [arXiv:1009.4657 [hep-th]].  


\bibitem{Bergshoeff:2011zk}
  E.~A.~Bergshoeff and F.~Riccioni,
  ``String Solitons and T-duality,''  JHEP {\bf 1105} (2011) 131  [arXiv:1102.0934 [hep-th]].  

\bibitem{Bergshoeff:2011qk}
  E.~A.~Bergshoeff and F.~Riccioni,
  ``The D-brane U-scan,''  arXiv:1109.1725 [hep-th].  

\bibitem{Bergshoeff:2012ex}
  E.~A.~Bergshoeff, A.~Marrani and F.~Riccioni,
  ``Brane orbits,''  Nucl.\ Phys.\ B {\bf 861} (2012) 104  [arXiv:1201.5819 [hep-th]].  

\bibitem{Houart:2011sk}
  L.~Houart, A.~Kleinschmidt and J.~Lindman Hornlund,
  ``An M-theory solution from null roots in $E_{11}$,''  JHEP {\bf 1101} (2011) 154  [arXiv:1101.2816 [hep-th]].  


\bibitem{Kleinschmidt:2011vu}
  A.~Kleinschmidt,
  ``Counting supersymmetric branes,''  JHEP {\bf 1110} (2011) 144  [arXiv:1109.2025 [hep-th]].  




\bibitem{Bergshoeff:1989qh}
  E.~A.~Bergshoeff, R.~E.~Kallosh and M.~Rakhmanov,
  ``Singlets Of Fermionic Gauge Symmetries,''  Phys.\ Lett.\ B {\bf 223} (1989) 391.  

\bibitem{Atick:1985iy}
  J.~J.~Atick, A.~Dhar and B.~Ratra,
  ``Superstring Propagation In Curved Superspace In The Presence Of Background Superyang-mills Fields,''  Phys.\ Lett.\ B {\bf 169} (1986) 54.  



\bibitem{Townsend:1983xt}
  P.~K.~Townsend,
  ``A New Anomaly Free Chiral Supergravity Theory From Compactification On K3,''  Phys.\ Lett.\ B {\bf 139} (1984) 283.  

\bibitem{Kleinschmidt:2004dy}
  A.~Kleinschmidt and H.~Nicolai,
  ``E(10) and SO(9,9) invariant supergravity,''  JHEP {\bf 0407} (2004) 041  [hep-th/0407101].  





\bibitem{Bergshoeff:2011mh}
  E.~A.~Bergshoeff and F.~Riccioni,
  ``Dual doubled geometry,''  Phys.\ Lett.\ B {\bf 702} (2011) 281  [arXiv:1106.0212 [hep-th]].  



\bibitem{Bergshoeff:2011se}
  E.~Bergshoeff, T.~Ort\'\i n and F.~Riccioni,
  ``Defect Branes,''  Nucl.\ Phys.\ B {\bf 856} (2012) 210  [arXiv:1109.4484 [hep-th]].  


\bibitem{Hull:1994ys}
  C.~M.~Hull and P.~K.~Townsend,
  ``Unity of superstring dualities,''  Nucl.\ Phys.\ B {\bf 438} (1995) 109  [hep-th/9410167].  





\bibitem{VanProeyen:1999ni}
  A.~Van Proeyen, ``Tools for supersymmetry,'' lectures given at the
Spring School on Quantum Field Theory: Supersymmetry and Superstrings,
24-30 Apr 1998, Calimanesti, Romania,
    hep-th/9910030.  



\bibitem{Hull}
  C.~M.~Hull,
  ``A Geometry for non-geometric string backgrounds,''  JHEP {\bf 0510} (2005) 065  [hep-th/0406102];  
  C.~M.~Hull,
  ``Doubled Geometry and T-Folds,''  JHEP {\bf 0707}, 080 (2007)  [hep-th/0605149];  
  C.~M.~Hull and R.~A.~Reid-Edwards,
  ``Gauge symmetry, T-duality and doubled geometry,''  JHEP {\bf 0808} (2008) 043  [arXiv:0711.4818 [hep-th]].  



\bibitem{Hull:2009mi}
  C.~Hull and B.~Zwiebach,
  ``Double Field Theory,''  JHEP {\bf 0909} (2009) 099  [arXiv:0904.4664 [hep-th]].  

\bibitem{Sagnotti:1987tw}
  A.~Sagnotti,
  ``Open Strings and their Symmetry Groups,''  in Cargese 1987, proceedings, nonperturbative quantum field theory,  521-528 and Rome II
univ. - ROM2F-87-025 (87,Rec.Mar.88) 12p  [hep-th/0208020].  





\bibitem{Henneaux:2007ej}
  M.~Henneaux, D.~Persson and P.~Spindel,
  ``Spacelike Singularities and Hidden Symmetries of Gravity,''  Living Rev.\ Rel.\  {\bf 11} (2008) 1  [arXiv:0710.1818 [hep-th]].  

\bibitem{Riccioni:2008jz}
  F.~Riccioni, A.~Van Proeyen and P.~C.~West,
  ``Real forms of very extended Kac-Moody algebras and theories with eight supersymmetries,''  JHEP {\bf 0805} (2008) 079  [arXiv:0801.2763 [hep-th]].


\bibitem{Bergshoeff:2007qi}
  E.~A.~Bergshoeff, I.~De Baetselier and T.~A.~Nutma,
  ``E(11) and the embedding tensor,''  JHEP {\bf 0709} (2007) 047  [arXiv:0705.1304 [hep-th]].  




\end{thebibliography}
\end{document}